\documentclass[pra,preprint,superscriptaddress,floatfix,showkeys,showpacs]{revtex4-1}

\usepackage{amsmath,amsfonts,amssymb}
\usepackage{graphicx,color}
\allowdisplaybreaks

\def\beq{\begin{equation}}
\def\eeq{\end{equation}}
\def\beqn{\begin{eqnarray}}
\def\eeqn{\end{eqnarray}}
\def\beqs{\begin{subequations}}
\def\eeqs{\end{subequations}}

\def\r {{\bf r}}

\def\Q {{\bf Q}}

\def\r {{\bf r}}

\def\Q {{\bf Q}}

\begin{document}

\title{Morphology of an interacting three-dimensional trapped Bose-Einstein
condensate from many-particle variance anisotropy}
\author{Ofir E. Alon}
\email{ofir@research.haifa.ac.il}
\affiliation{Department of Mathematics, University of Haifa, Haifa 3498838, Israel}
\affiliation{Haifa Research Center for Theoretical Physics and Astrophysics, University of Haifa,
Haifa 3498838, Israel}

\begin{abstract}
The variance of the position operator is associated with how wide or narrow a wave-packet is,
the momentum variance is similarly correlated with the size of a wave-packet in momentum space,
and the angular-momentum variance quantifies to what extent a wave-packet is non-spherically symmetric.
We examine an interacting three-dimensional trapped Bose-Einstein condensate at the limit of
an infinite number of particles,
and investigate its position, momentum, and angular-momentum anisotropies.
Computing the variances of the three Cartesian components of the position, momentum, and angular-momentum operators
we present simple scenarios where the anisotropy of a Bose-Einstein condensate
is different at the many-body and mean-field levels of theory,
despite having the same many-body and mean-field densities per particle.
This suggests a way to classify correlations 
via the morphology of 100\% condensed bosons in a three-dimensional trap 
at the limit of an infinite number of particles.
Implications are briefly discussed.
\end{abstract}
\maketitle

\section{Introduction}\label{introduction}

There has been an increasing interest in the theory and properties of trapped
Bose-Einstein condensates at the limit of an infinite number of particles \cite{CASTIN,LIEB1,LIEB2,ERDOS1,ERDOS2,VAR1,VAR2,CEDERBAUM1,
MIX1,MIX2,MIX3,CEDERBAUM2}.
Here, one may divide the research questions into two, inter-connected groups.
The first group of research questions deals with rigorous results,
mainly proving when many-body and mean-field,
Gross-Pitaevskii theories coincide at this limit,
whereas the second group of questions deals with characterizing correlations
in a trapped Bose-Einstein condensate
based on the difference between many-body and mean-field
properties at the infinite-particle-number limit.
In \cite{LIEB1},
it has been shown that the ground-state energy per particle and density per particle
computed at the many-body level of theory coincide with the respective mean-field results.
The infinite-particle-number limit is defined
such that the interaction parameter, i.e., the product of the number of bosons times the interaction strength 
(proportional to the scattering length), is held constant.
Similarly, in \cite{LIEB2} it has been shown that the reduced one-body and any reduced finite-$n$-body density matrix \cite{RDMB1,RDMB2} per particle are 100\% condensed,
and that the leading natural orbital boils down to the Gross-Pitaevskii single-particle function.
Analogous results connecting time-dependent many-body and mean-field theories are given in \cite{ERDOS1,ERDOS2}
and developments for mixtures in \cite{MIX1,MIX2,MIX3}.
In \cite{CEDERBAUM2},
the many-boson wave-function at the infinite-particle-number limit has been constructed explicitly.

The difference between many-body and mean-field theories at the limit of an infinite number of particles,
which as stated above coincide at the level of the energy, densities, and reduced density matrices per particle,
starts to show up in variances of many-particle observables \cite{VAR1,VAR2}.
Of course, the wave-functions
themselves differ and their overlap is smaller than one \cite{CEDERBAUM1}.
In evaluating the variances of many-particle observables
two-body operators emerge whose combination with
the elements of the reduced two-body density matrix can pick up even
the tiniest depletion, which always exist due to the inter-particle interaction \cite{VAR1,VAR2}.
Here, the quantitative difference between the many-body and mean-field variances
is a useful tool to benchmark many-body numerical approaches \cite{RMP},
whereas the qualitative differences serve to define and characterize the nature
of correlations in 100\% condensed bosons at the infinite-particle-number limit.

Qualitative differences between the many-body and mean-field variances per particle
depend on both the system and observable under investigation
and emerge because the 100\% condensed bosons are interacting. 
In the ground-state of a one-dimensional double-well potential,
the mean-field position variance per particle increases monotonously with the interaction parameter
whereas, once about a single particle is excited outside the condensed mode,
the many-body position variance per particle starts to decrease \cite{VAR1}.
In the analogous time-dependent setup of a bosonic Josephson junction,
the mean-field variance is oscillating and bound by the size of the junction,
whereas the many-body variance increases to `sizes' several times larger \cite{VAR2}.
In two spatial dimensions additional features come out.
The position variance per particle in a thin annulus can exhibit a different dimensionality \cite{VAR3}
and both the position and momentum variances can exhibit opposite anisotropies \cite{VAR4}
when computed at the many-body and mean-field levels of theory
in an out-of-equilibrium quench dynamics.
In two spatial dimensions the many-particle variance
of the $\hat L_Z$ component of the angular-momentum operator becomes available,
and used to analyze the lack of conservation of symmetries in the mean-field dynamics \cite{VAR5}.

In the present work we analyze the many-particle position, momentum, and angular-momentum
variances of a three-dimensional anisotropic trapped Bose-Einstein condensate at
the limit of an infinite number of particles,
focusing on three-dimensional scenarios that do not have (one-dimensional and) two-dimensional analogs.
Mainly, the available permutations between the three Cartesian components of a many-particle operator,
such as the position and momentum operators,
allow one for various different
anisotropies of the respective mean-field and many-body variances than in two spatial dimensions \cite{VAR4}.
Furthermore,
anisotropy of the angular-momentum variance can only be investigated when
there is more than one component,
and this occurs with the three Cartesian components of the angular-momentum operator
in three spatial dimensions.

The structure of the paper is as follows.
In Sec.~\ref{theory} theory and definitions are developed.
In Sec.~\ref{applications} we present two applications where a common methodological line of investigation
is that the variances at the many-body level of theory can be computed analytically.
In Subsec.~\ref{position}, the anisotropy of the position and momentum variances in the out-of-equilibrium breathing dynamics of a Bose-Einstein condensate in a three-dimensional anisotropic harmonic
potential are analyzed,
and in Subsec.~\ref{angular},
a solvable model is devised which allows one to analyze 
the anisotropy of the angular-momentum variances in the ground state
of interacting bosons in a three-dimensional anisotropic harmonic potential.
All quantities are computed at the infinite-particle-number limit.
Finally, we summarize in Sec.~\ref{summary}.

\section{Theory}\label{theory}

The variances per particle of a many-particle 
observable $\hat O = \sum_j^N \hat o_j$
computed at the many-body (MB) and mean-field, Gross-Pitaevskii (GP) levels of theory 
are connected at the
limit of an infinite number of particles
by the following relation \cite{VAR1}:
\beq\label{VAR_INF}
{\rm Var|_{MB}}(\hat O) = {\rm Var|_{GP}}(\hat O) + {\rm Var|_{correlations}}(\hat O).
\eeq
Here, ${\rm Var|_{MB}}(\hat O)=\lim_{N \to \infty}
\left[\frac{1}{N}(\langle\Psi|\hat O^2|\Psi\rangle-\langle\Psi|\hat O|\Psi\rangle^2)\right]$,
where $\Psi$ is the solution of the many-particle Schr\"odinger equation, 
and
${\rm Var|_{GP}}(\hat O) = \langle\psi_{GP}|\hat o^2|\psi_{GP}\rangle-\langle\psi_{GP}|\hat o|\psi_{GP}\rangle^2$,
where $\psi_{GP}$ is the solution of the corresponding Gross-Pitaevskii equation.
Recall that the infinite-particle-number limit is defined 
such that the interaction parameter, i.e., the product of the number of bosons times the interaction strength, is kept fixed. 
The correlations term, ${\rm Var|_{correlations}}(\hat O)$, quantifying the difference between the mean-field and many-body variances, depends on the elements of the reduced two-body density matrix 
where at least one of the indexes corresponds to a natural orbital higher than the condensed mode
\cite{VAR1}.
For non-interacting bosons the correlations term obviously vanishes.
As stated above,
one is interested in qualitative differences between ${\rm Var|_{GP}}(\hat O)$ and ${\rm Var|_{MB}}(\hat O)$
and their origin.

Consider a Bose-Einstein condensate for which the many-body variances of the three Cartesian components
of, say, the position operator are different and satisfy, without loss of generality,
the anisotropy
\beqs\label{VAR_ANI_3D}
\beq\label{VAR_ANI_3D_MB}
{\rm Var|_{MB}}(\hat X) > {\rm Var|_{MB}}(\hat Y) > {\rm Var|_{MB}}(\hat Z).
\eeq
We define the following classification with respect to the possible different anisotropies of the
respective mean-field position variances:
\beqn\label{VAR_ANI_3D_GP}
\{1\}: \quad & & \Big\{ {\rm Var|_{GP}}(\hat X) > {\rm Var|_{GP}}(\hat Y) > {\rm Var|_{GP}}(\hat Z) \Big\}, \nonumber \\
\{1,2\}: \quad & & \Big\{ {\rm Var|_{GP}}(\hat Y) > {\rm Var|_{GP}}(\hat X) > {\rm Var|_{GP}}(\hat Z); \nonumber \\
         \quad & & \ \ {\rm Var|_{GP}}(\hat X) > {\rm Var|_{GP}}(\hat Z) > {\rm Var|_{GP}}(\hat Y); \nonumber \\
         \quad & & \ \ {\rm Var|_{GP}}(\hat Z) > {\rm Var|_{GP}}(\hat Y) > {\rm Var|_{GP}}(\hat X) \Big\}, \nonumber \\
\{1,2,3\}: \quad & & \Big\{ {\rm Var|_{GP}}(\hat Y) > {\rm Var|_{GP}}(\hat Z) > {\rm Var|_{GP}}(\hat X); \nonumber \\
         \quad & & \ \ {\rm Var|_{GP}}(\hat Z) > {\rm Var|_{GP}}(\hat X) > {\rm Var|_{GP}}(\hat Y) \Big\}. \
\eeqn
\eeqs
Naturally, the classification (\ref{VAR_ANI_3D_GP}) follows the classes of the $S_3$ permutation group denoted by
$\{1\}$, $\{1,2\}$, and $\{1,2,3\}$.
If the mean-field variances exhibit anisotropy other than the anisotropy of the respective many-body variances,
i.e., the ordering of the former does not belong to the class $\{1\}$,
we may interpret that the mean-field and many-body morphologies 
of the Bose-Einstein condensate with respect to the operators under investigation are distinct.
In the present work we investigate manifestations of definition (\ref{VAR_ANI_3D}) 
utilizing the many-particle position ($\hat X, \hat Y, \hat Z$), momentum ($\hat P_X, \hat P_Y, \hat P_Z$),
and angular-momentum ($\hat L_X, \hat L_Y, \hat L_Z$) operators
for classifying the morphology of 100\% condensed trapped bosons at the limit of an infinite number of particles.

\section{Applications}\label{applications}

\subsection{Position and momentum variances in an out-of-equilibrium dynamics of a
three-dimensional trapped Bose-Einstein condensate}\label{position}

\begin{figure}[!]
\begin{center}
\vglue -2.0 truecm
\hglue -1.0 truecm
\includegraphics[width=0.345\columnwidth,angle=-90]{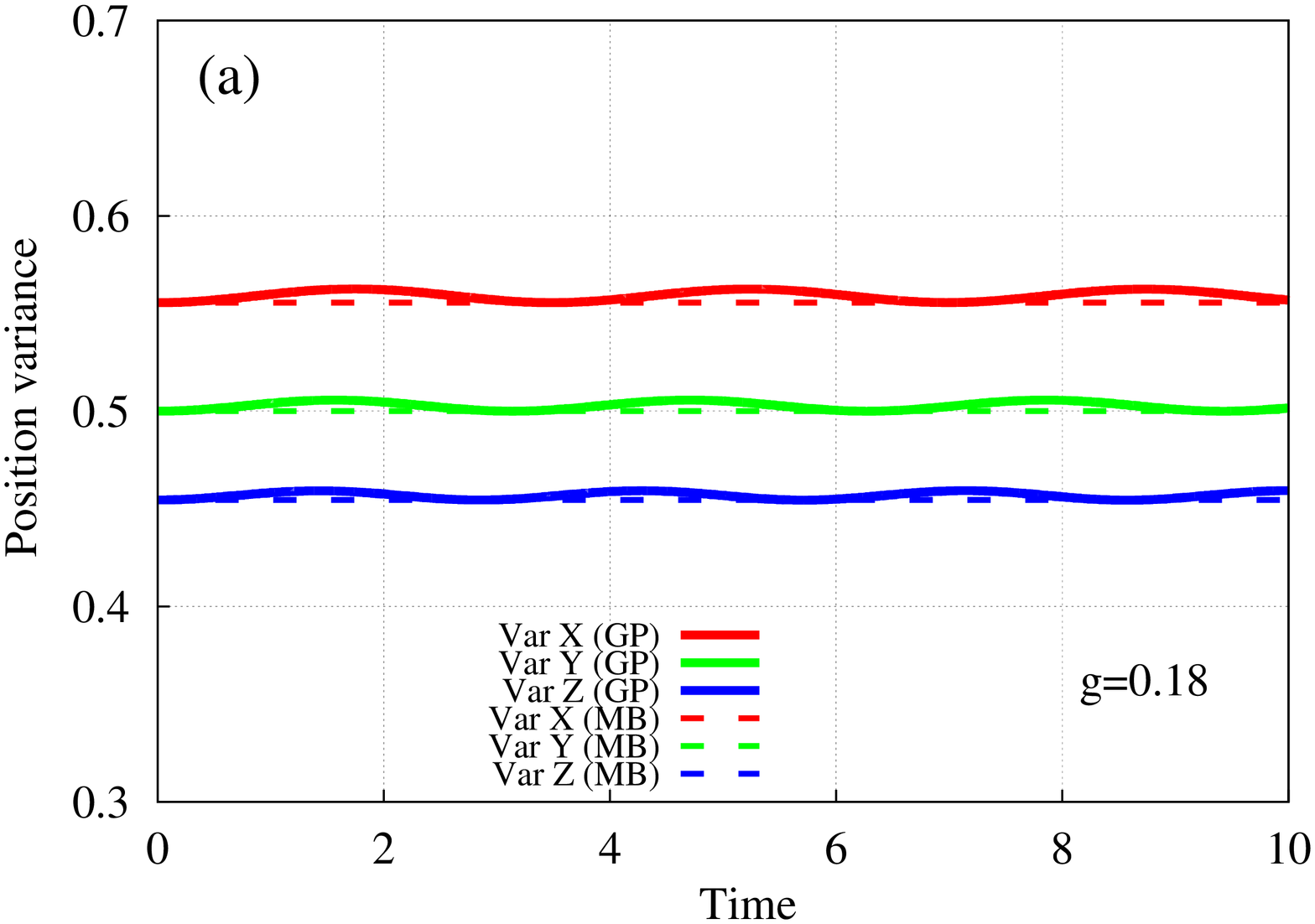}
\includegraphics[width=0.345\columnwidth,angle=-90]{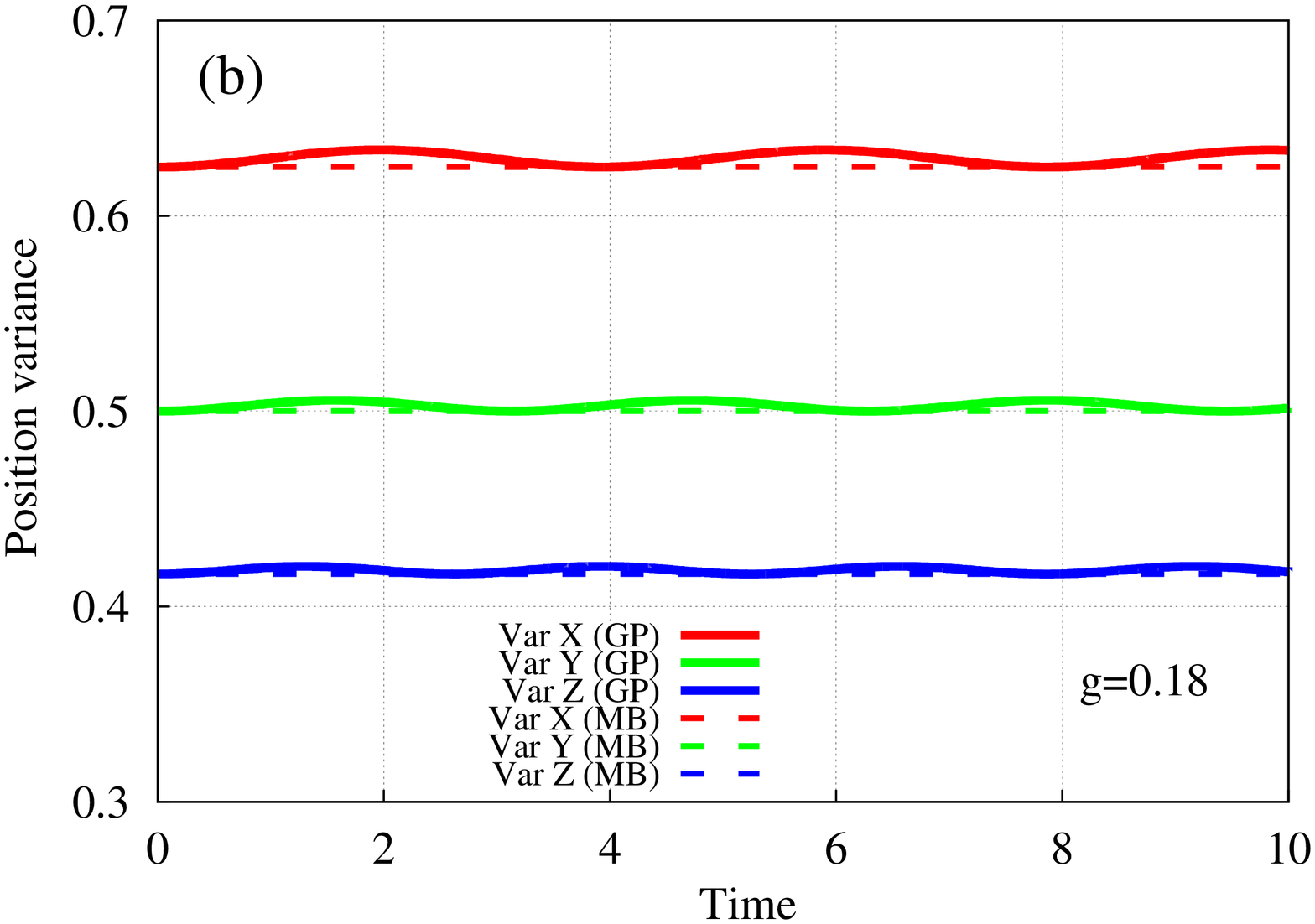}
\vglue -0.25 truecm
\hglue -1.0 truecm
\includegraphics[width=0.345\columnwidth,angle=-90]{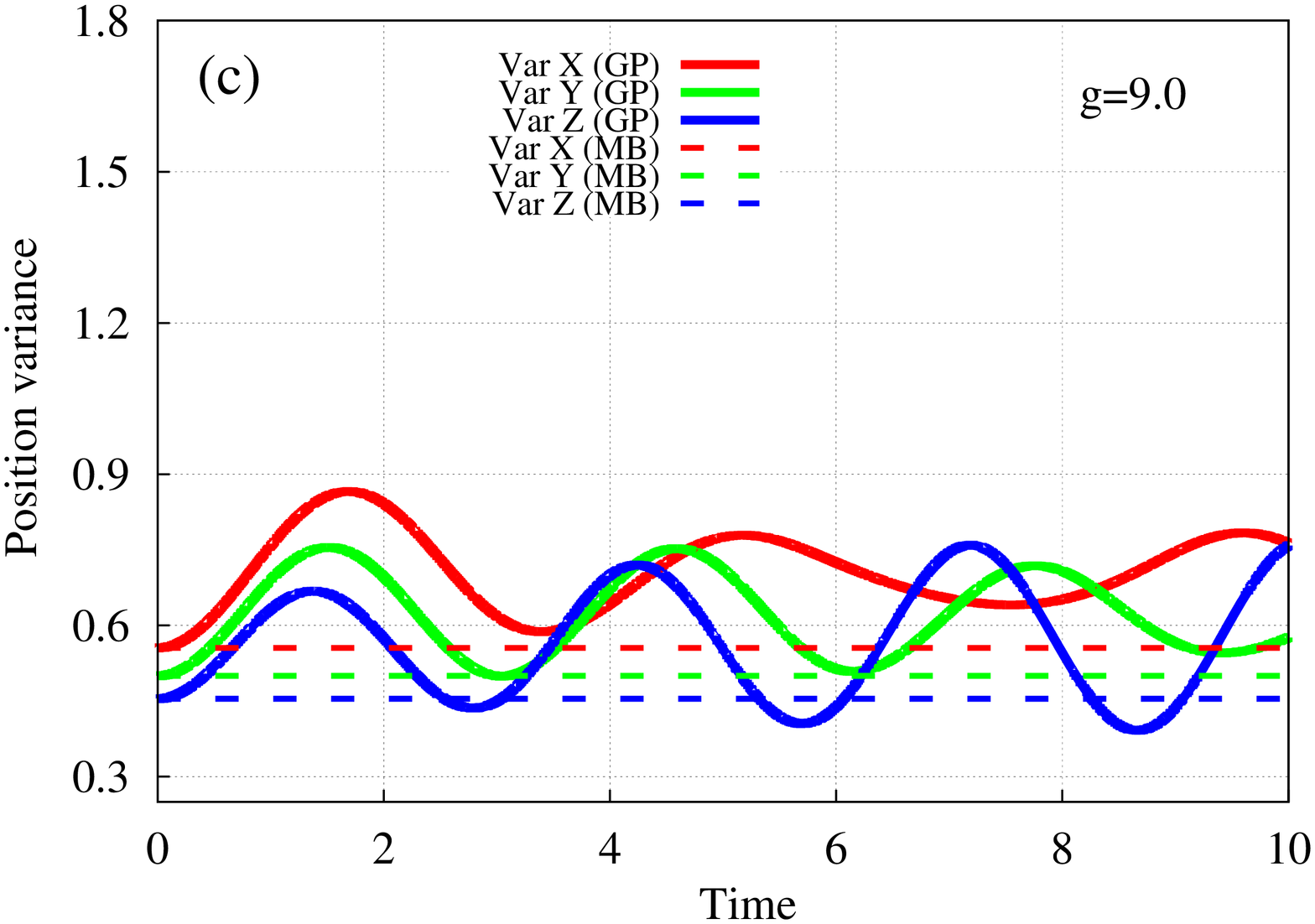}
\includegraphics[width=0.345\columnwidth,angle=-90]{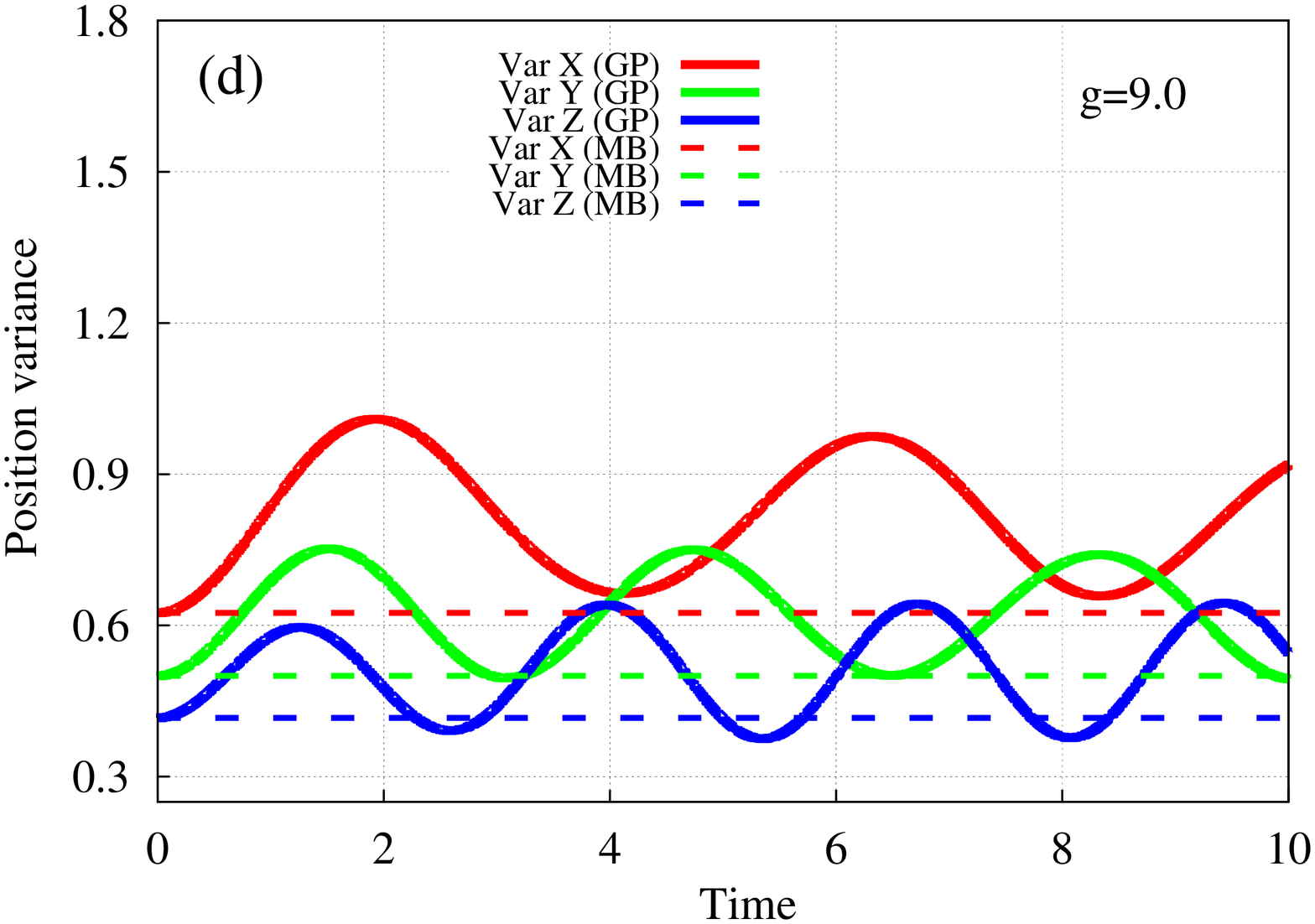}
\vglue -0.25 truecm
\hglue -1.0 truecm
\includegraphics[width=0.345\columnwidth,angle=-90]{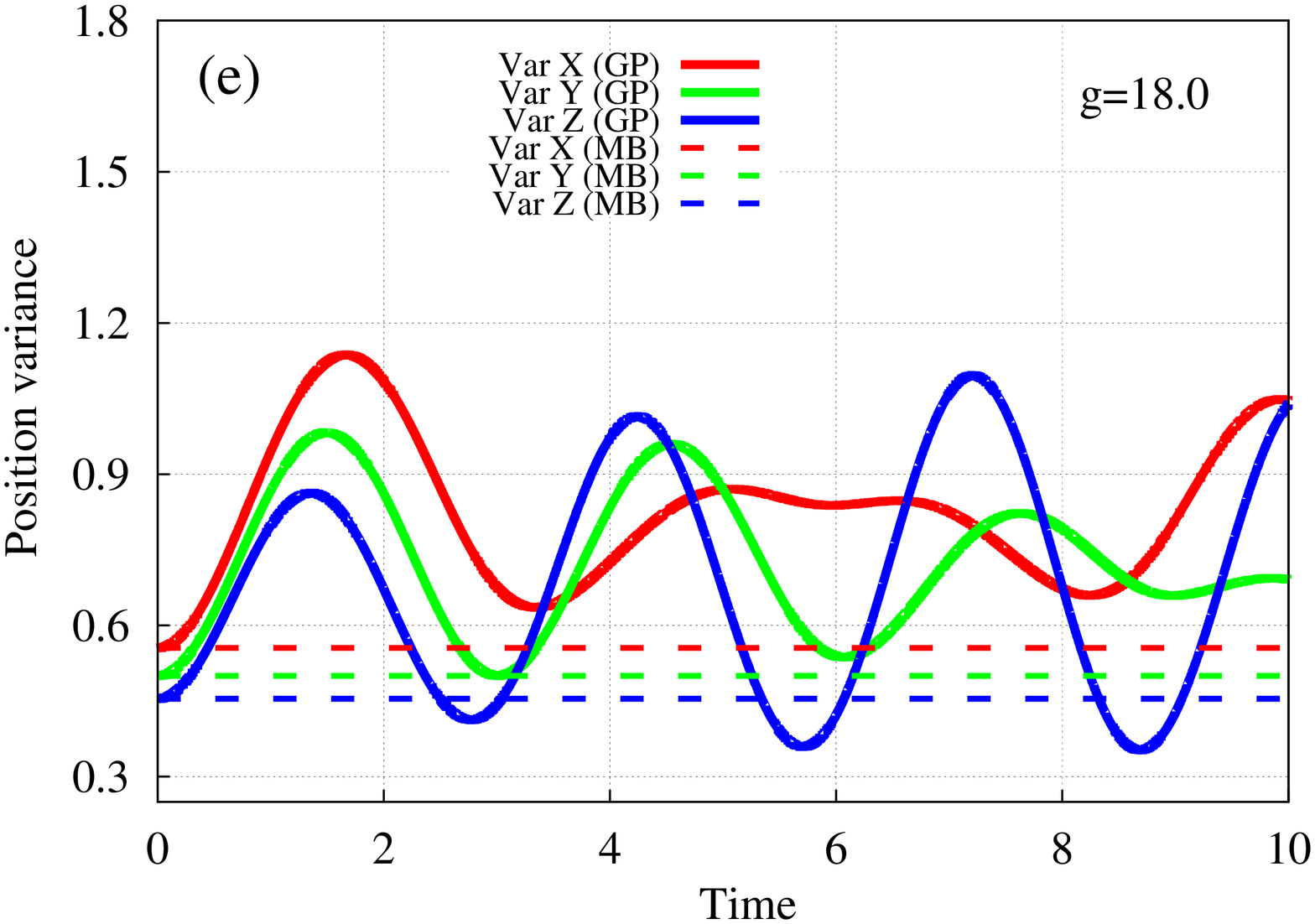}
\includegraphics[width=0.345\columnwidth,angle=-90]{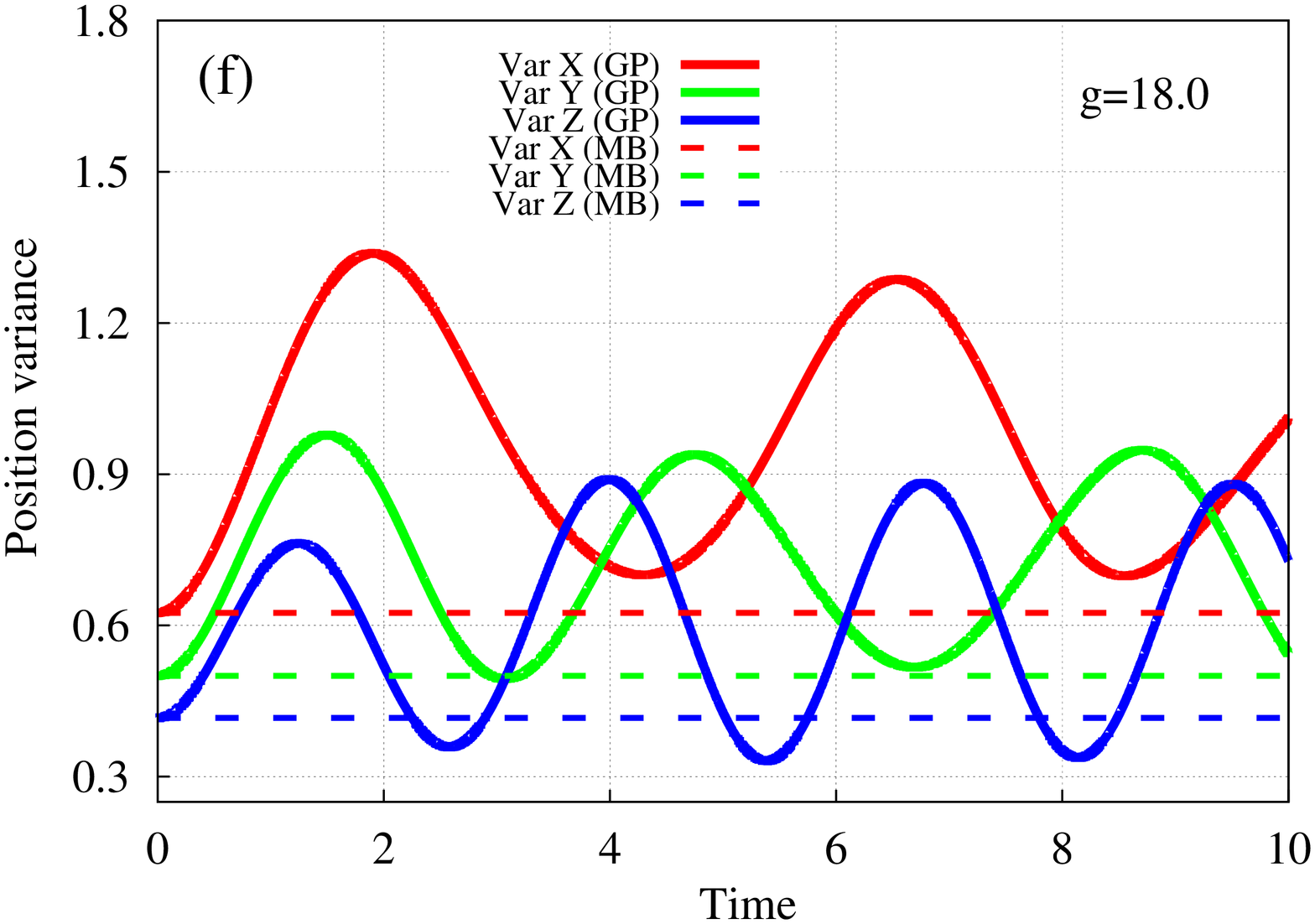}
\vglue -0.25 truecm
\hglue -1.0 truecm
\includegraphics[width=0.345\columnwidth,angle=-90]{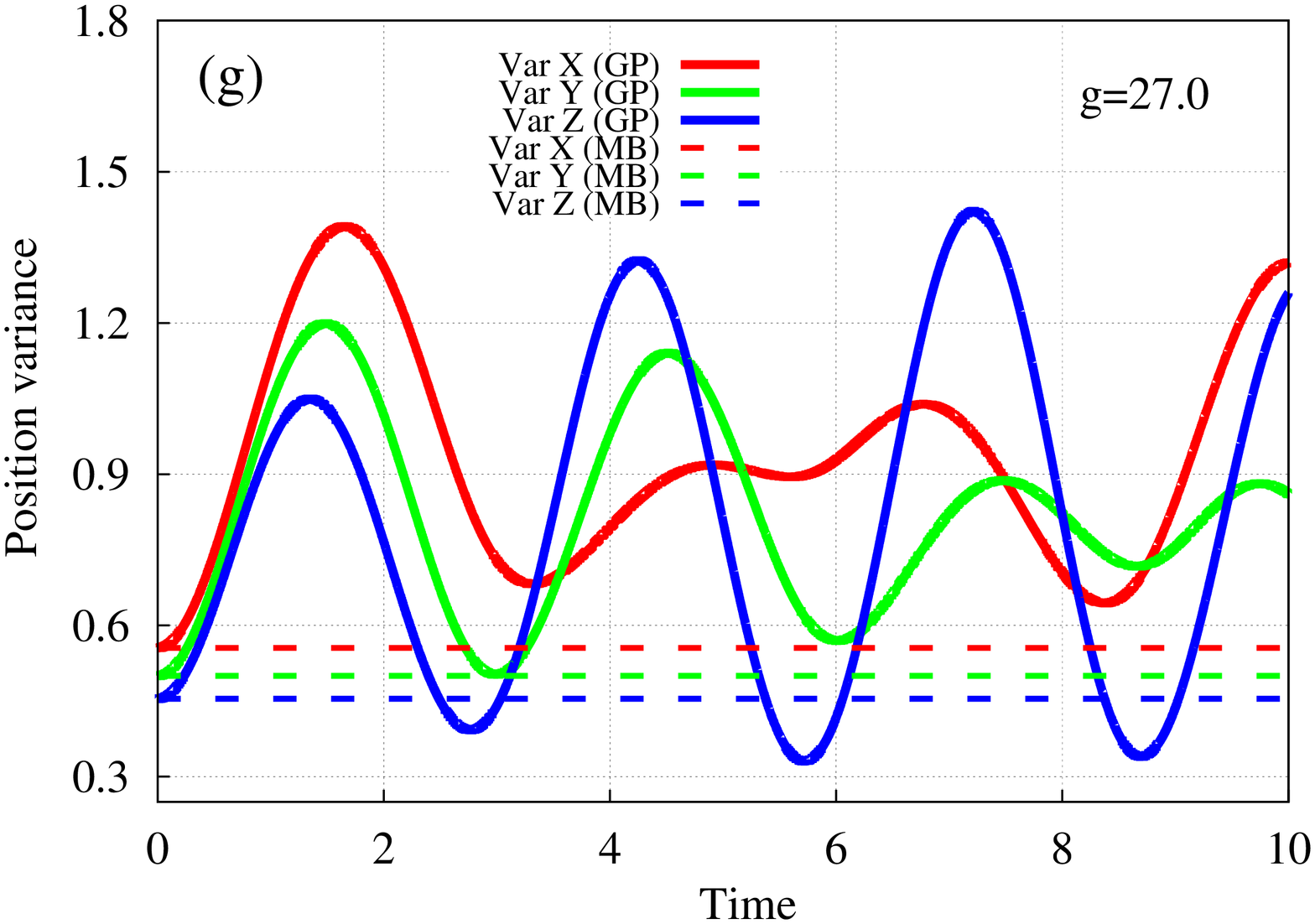}
\includegraphics[width=0.345\columnwidth,angle=-90]{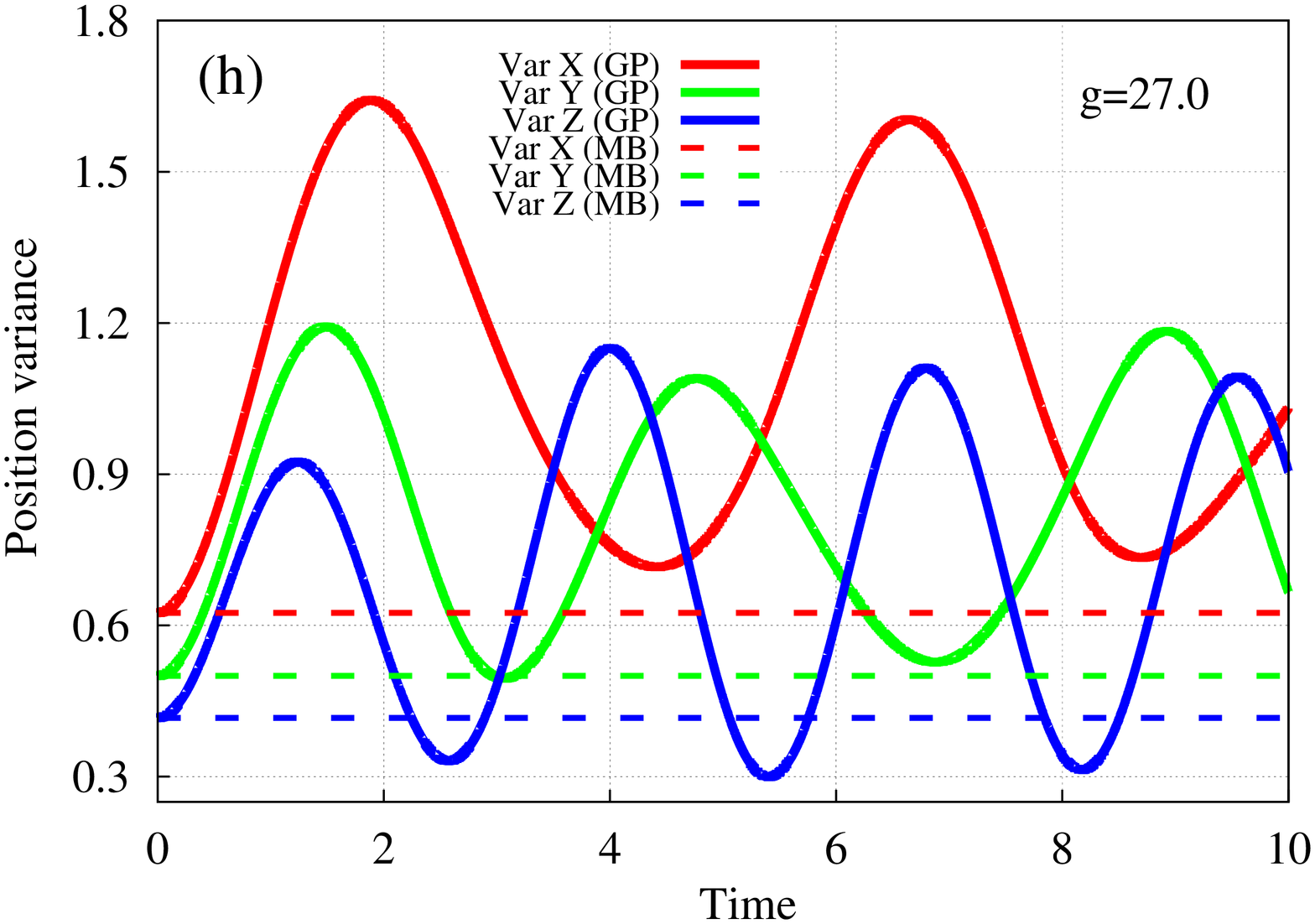}
\end{center}
\vglue -0.275 truecm
\caption{Many-particle position ($\hat X$, $\hat Y$, and $\hat Z$; in red, green, and blue)
variance per particle as a function of time 
computed at the limit of an infinite number of particles
within many-body (dashed lines) and mean-field (solid lines) levels of theory in an interaction-quench scenario.
The harmonic trap is 10\% anisotropic in panels (a), (c), (e), (g) 
and 20\% anisotropic in panels (b), (d), (f), (h).
The coupling constant $g$ is indicated in each panel.  
Different anisotropy classes of the position variance emerge with time.
See the text for more details.
The quantities shown are dimensionless.}
\label{f1}
\end{figure}

\begin{figure}[!]
\begin{center}
\vglue -2.0 truecm
\hglue -1.0 truecm
\includegraphics[width=0.345\columnwidth,angle=-90]{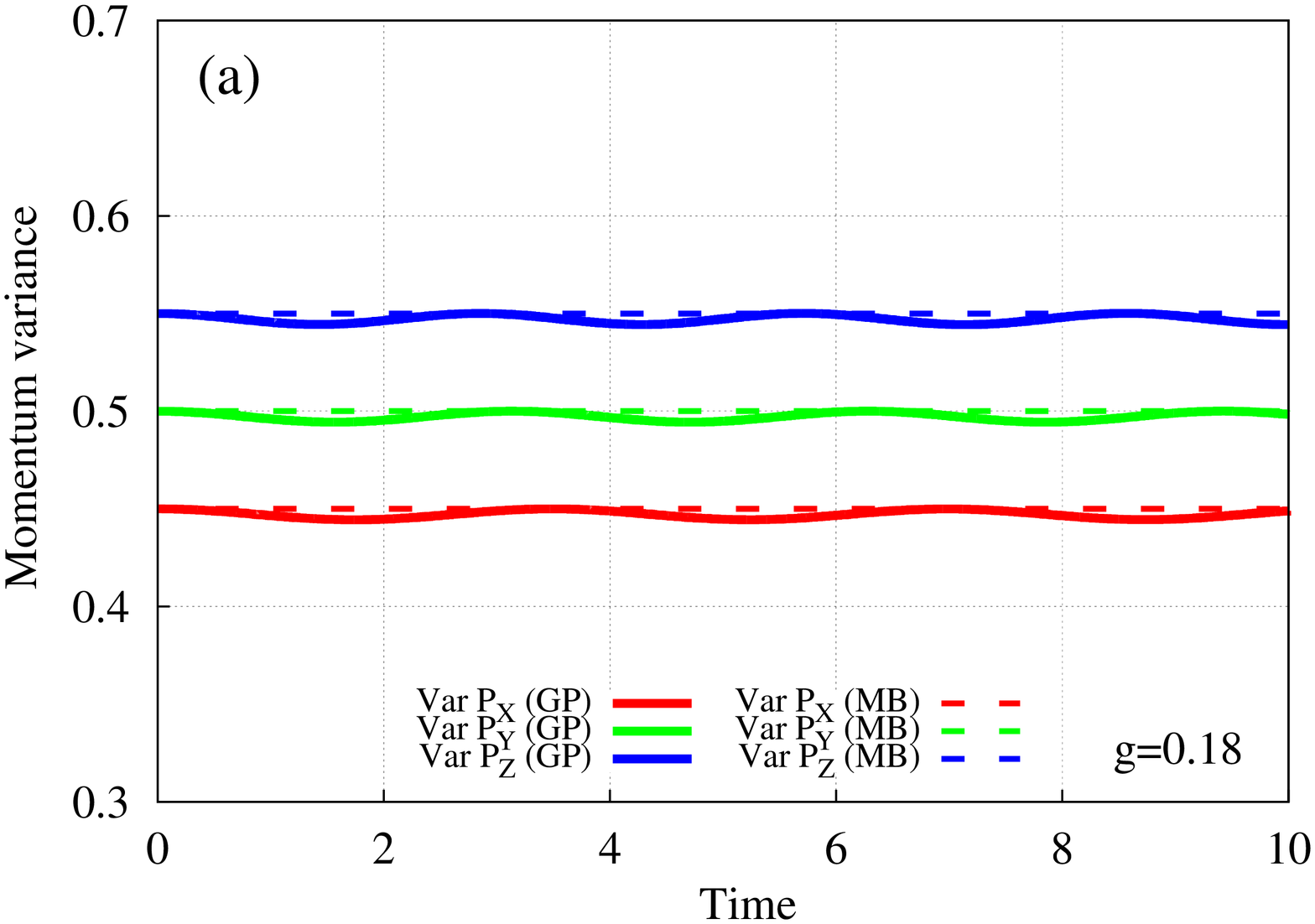}
\includegraphics[width=0.345\columnwidth,angle=-90]{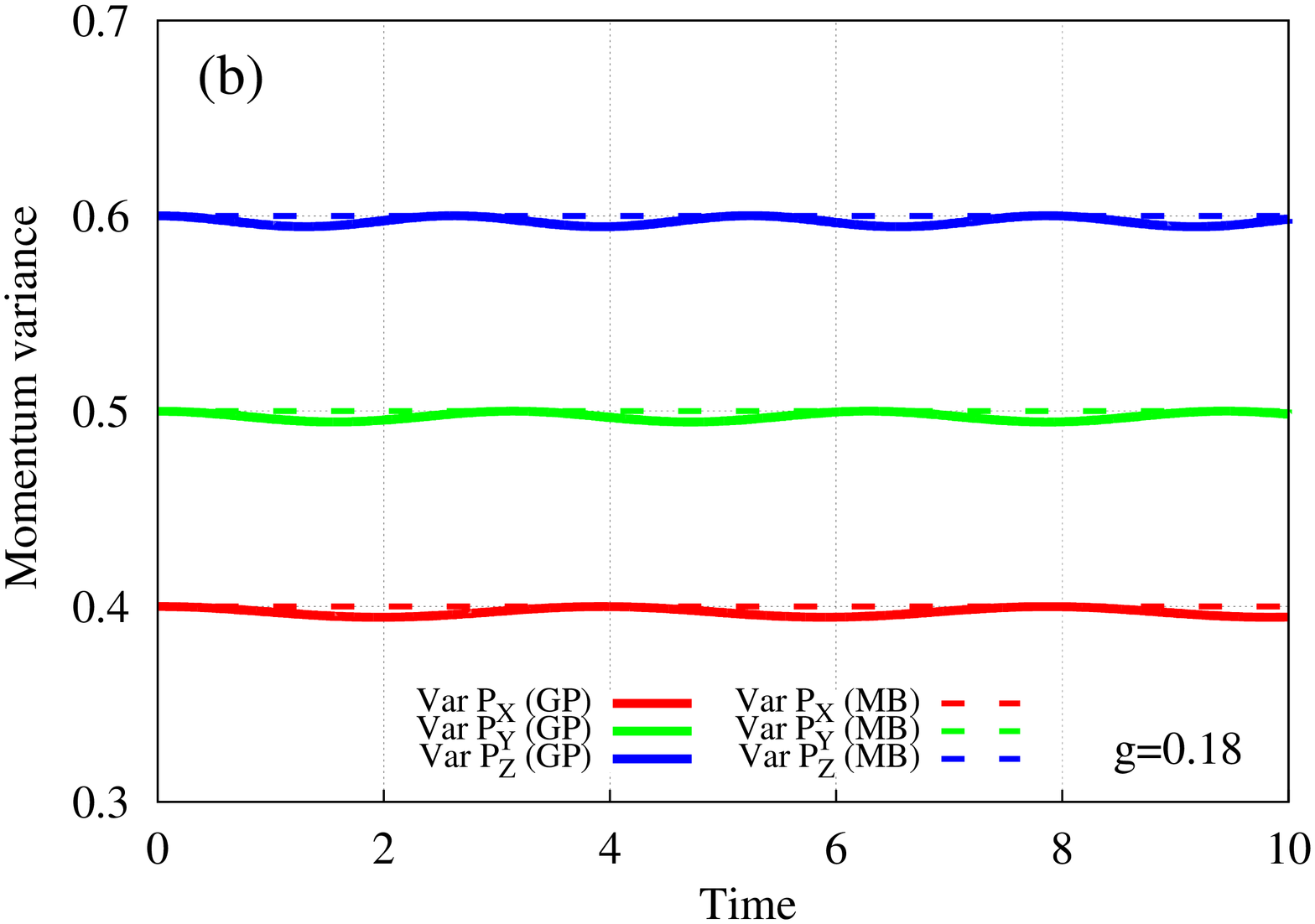}
\vglue -0.25 truecm
\hglue -1.0 truecm
\includegraphics[width=0.345\columnwidth,angle=-90]{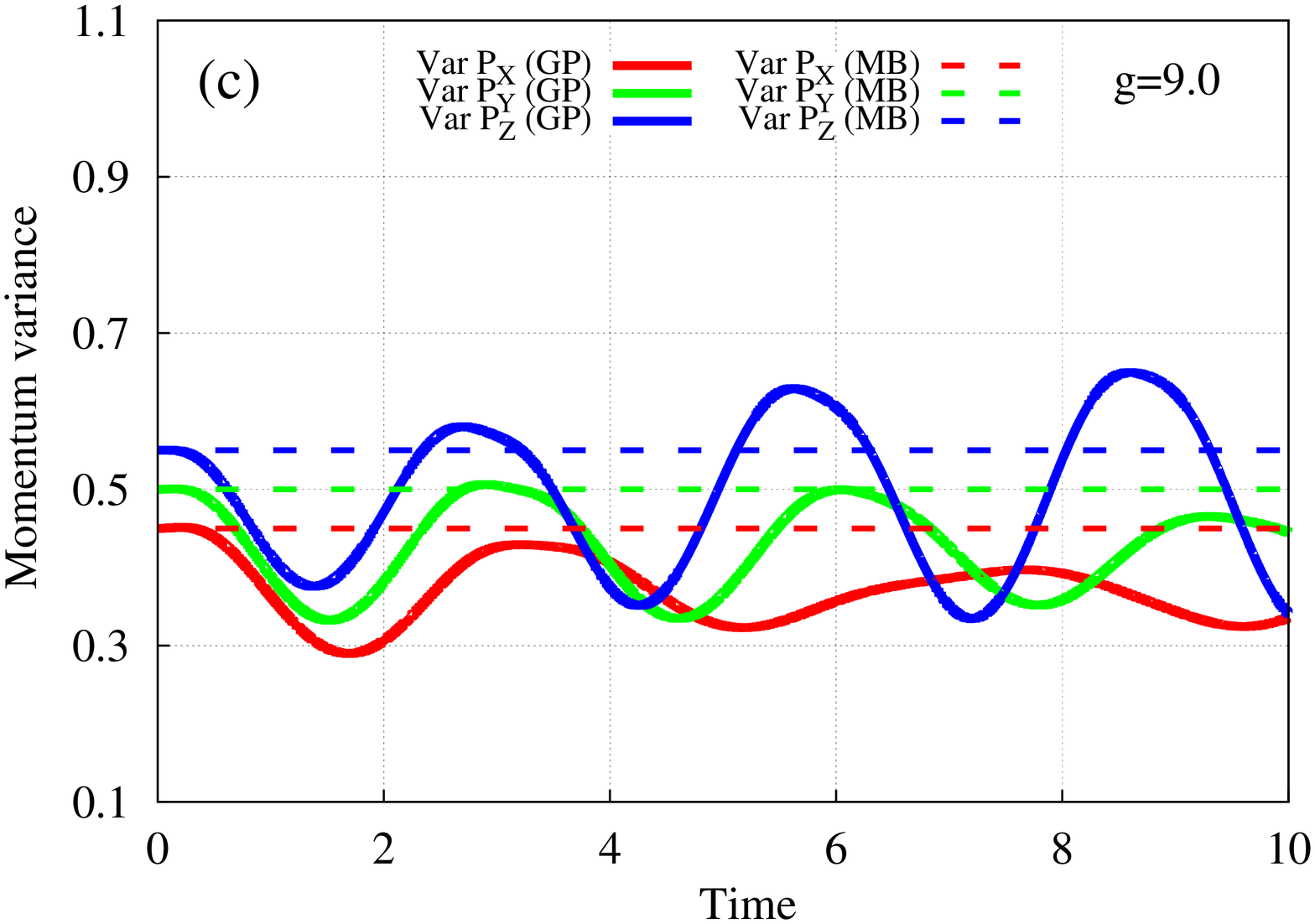}
\includegraphics[width=0.345\columnwidth,angle=-90]{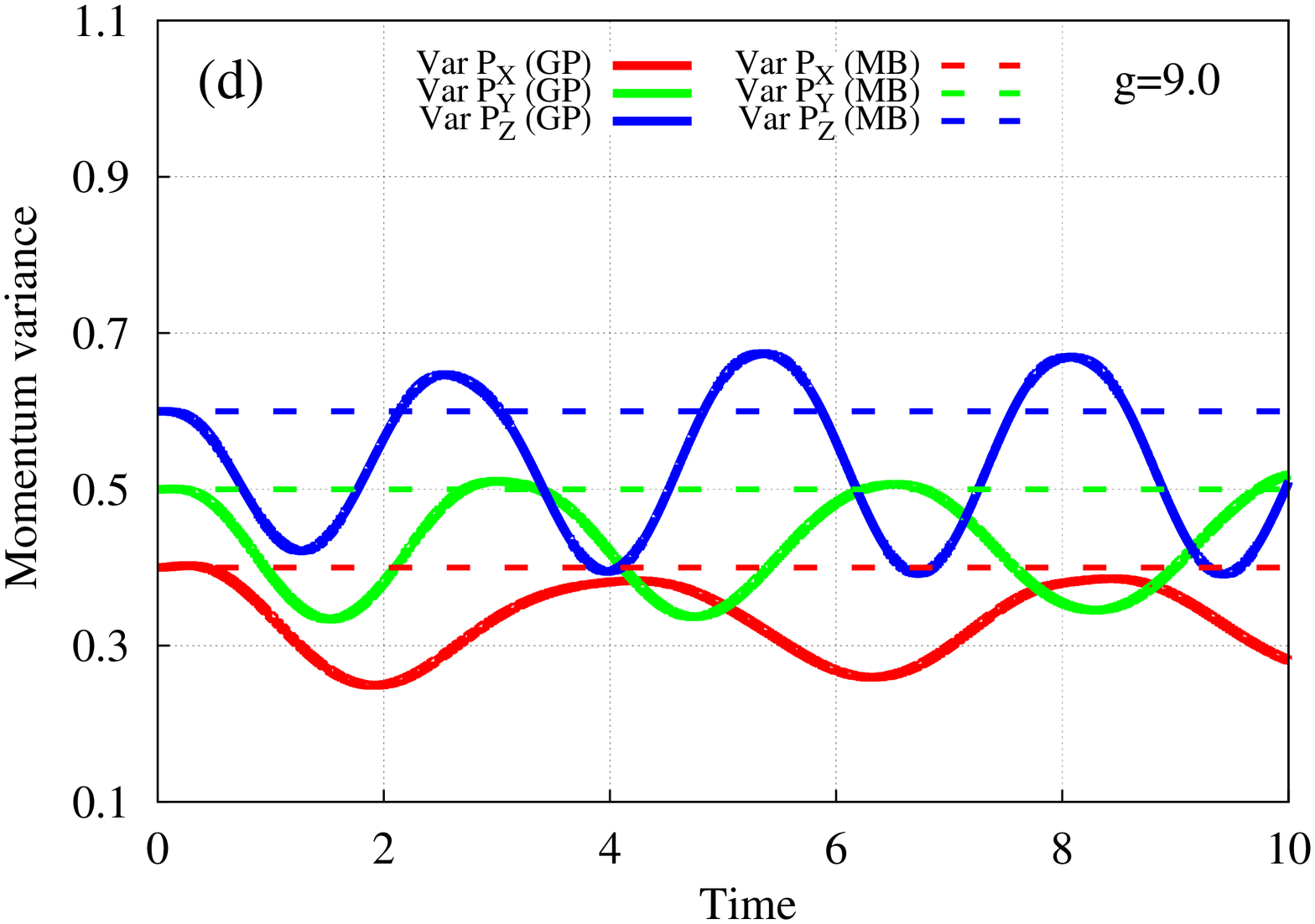}
\vglue -0.25 truecm
\hglue -1.0 truecm
\includegraphics[width=0.345\columnwidth,angle=-90]{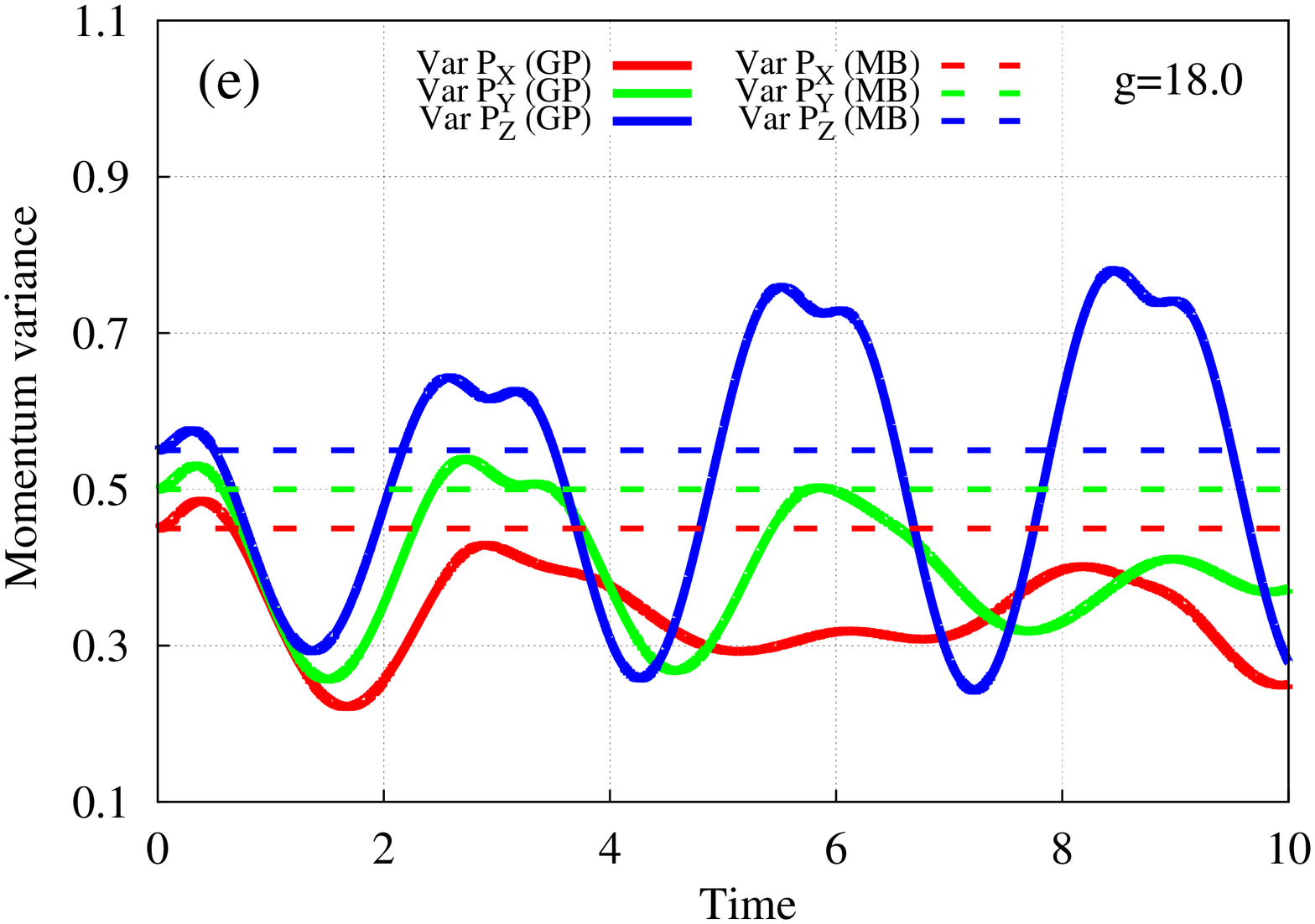}
\includegraphics[width=0.345\columnwidth,angle=-90]{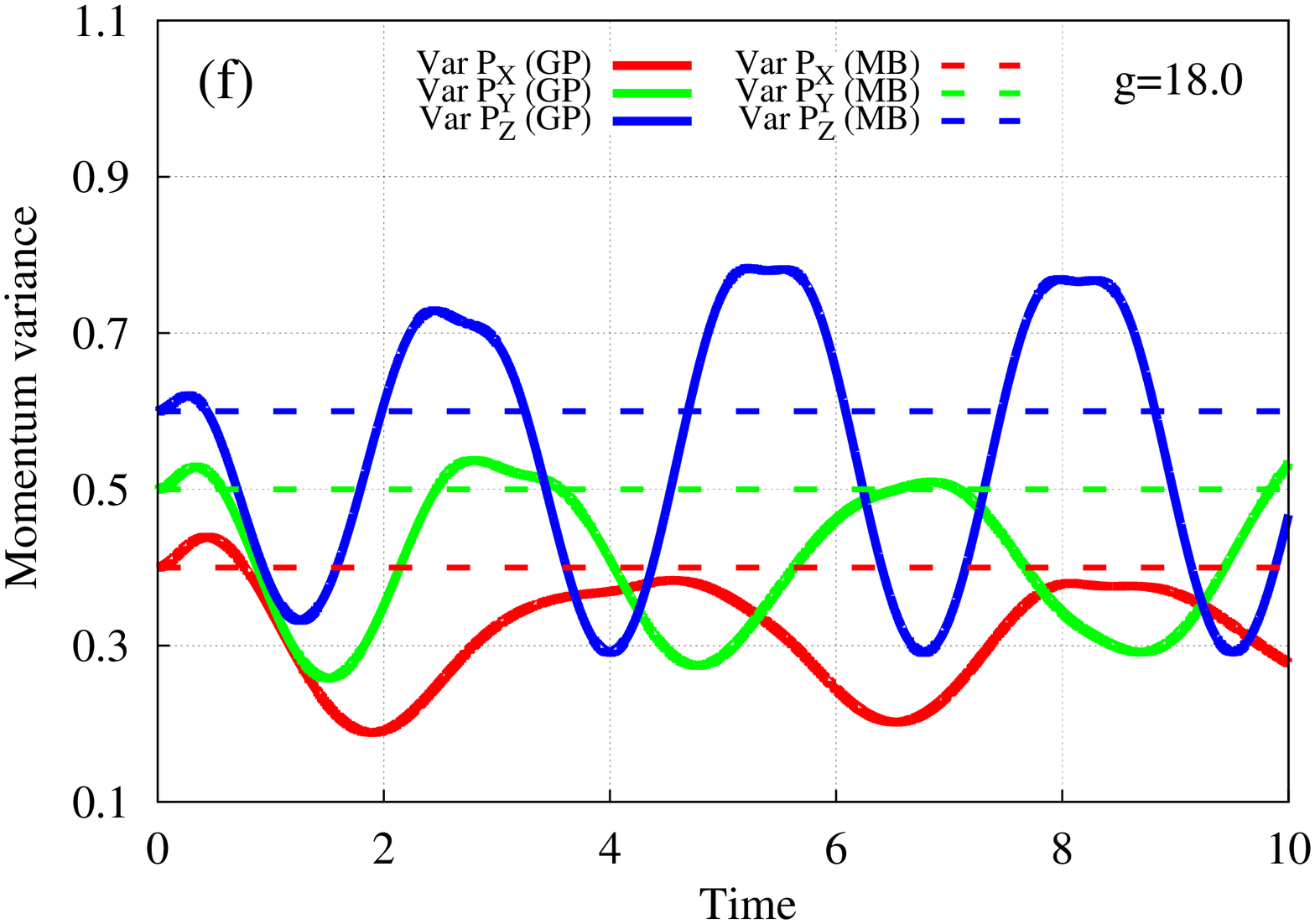}
\vglue -0.25 truecm
\hglue -1.0 truecm
\includegraphics[width=0.345\columnwidth,angle=-90]{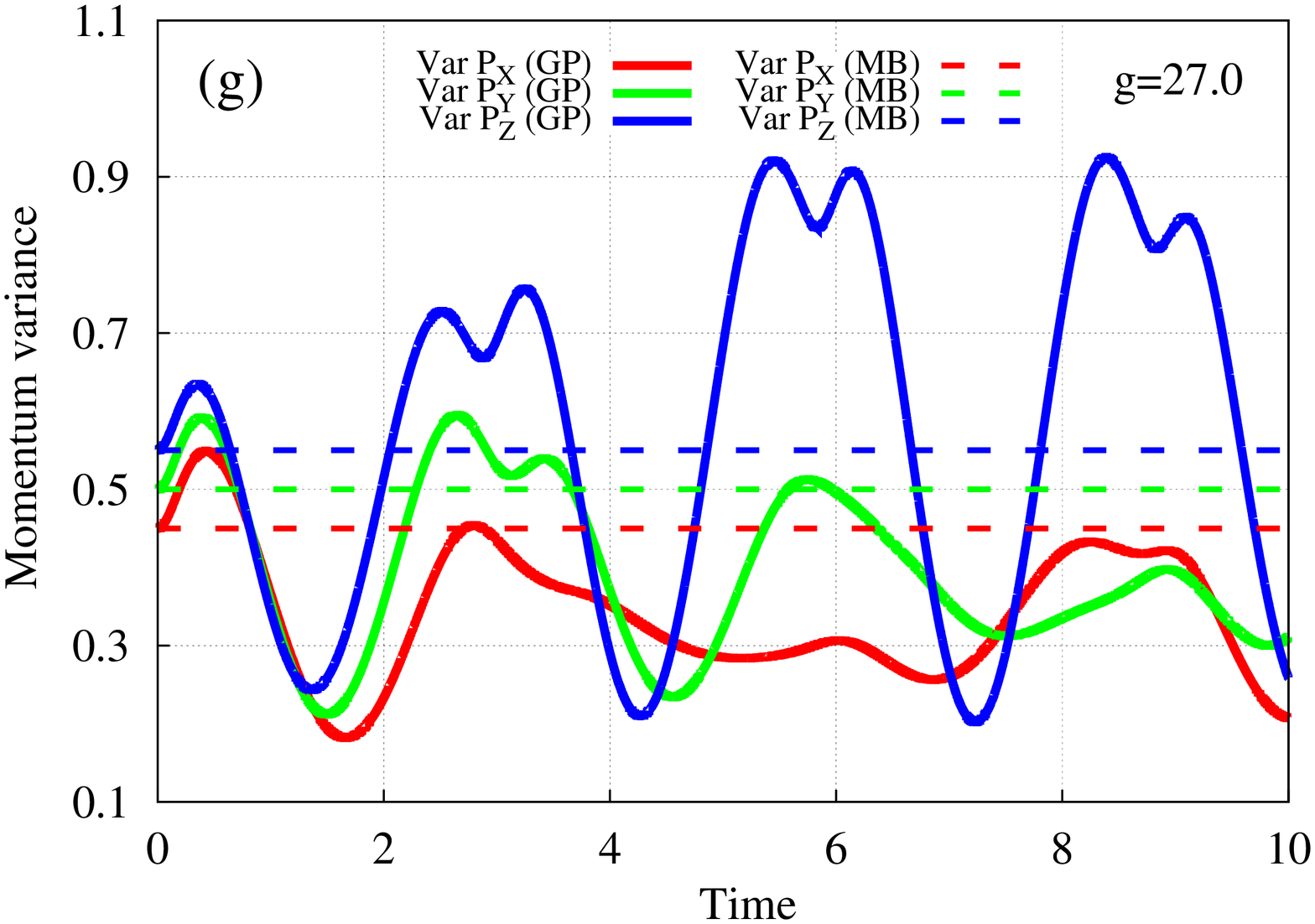}
\includegraphics[width=0.345\columnwidth,angle=-90]{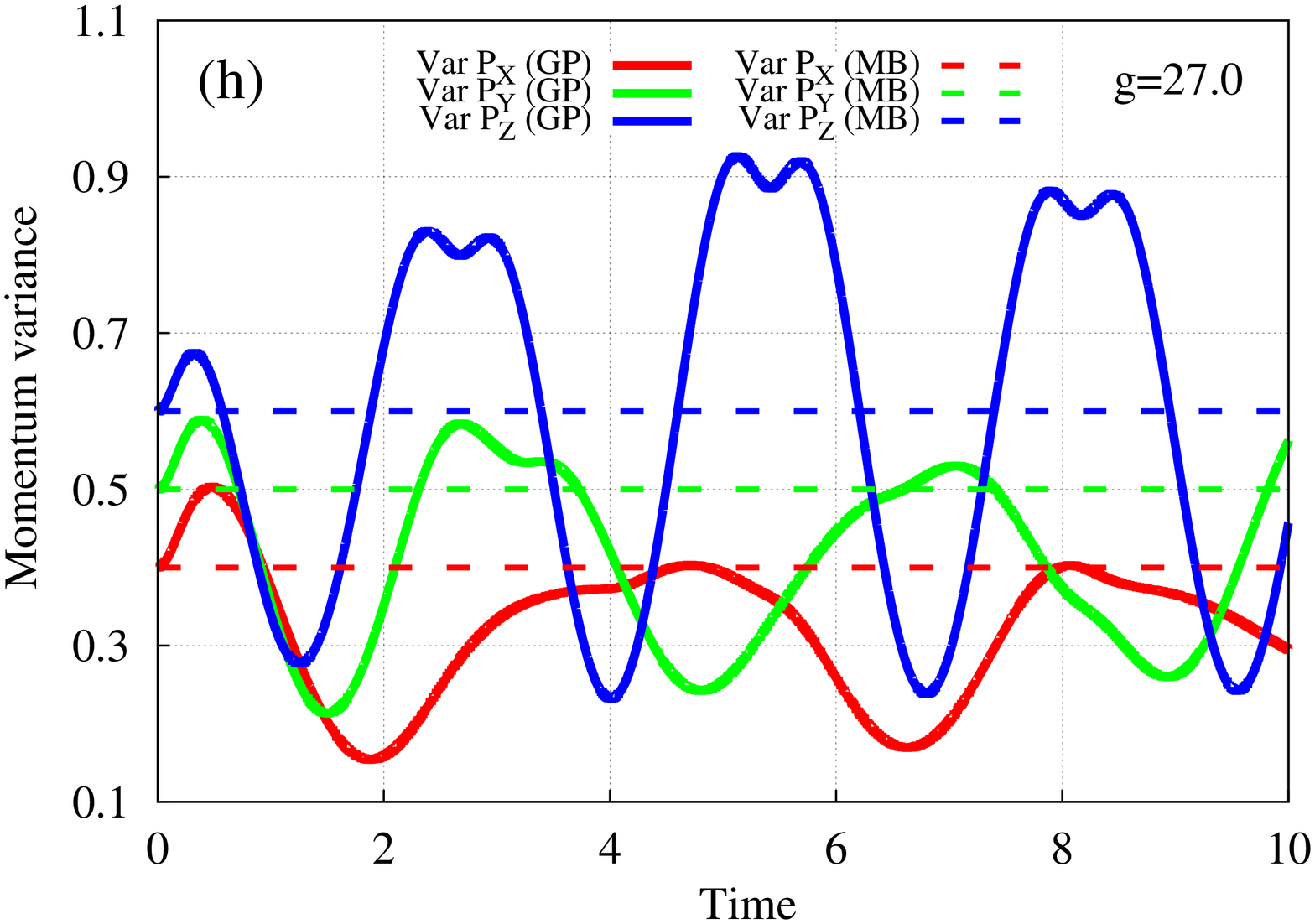}
\end{center}
\vglue -0.275 truecm
\caption{Many-particle momentum ($\hat P_X$, $\hat P_Y$, and $\hat P_Z$; in red, green, and blue)
variance per particle as a function of time 
computed at the infinite-particle-number limit
within many-body (dashed lines) and mean-field (solid lines) levels of theory in an interaction-quench scenario.
The harmonic trap is 10\% anisotropic in panels (a), (c), (e), (g) 
and 20\% anisotropic in panels (b), (d), (f), (h).
The coupling constant $g$ is indicated in each panel.  
Different anisotropy classes of the momentum variance emerge with time.
See the text for more details.
The quantities shown are dimensionless.}
\label{f2}
\end{figure}

Consider $N$ structureless bosons trapped in a three-dimensional anisotropic harmonic potential
and interacting by a general two-body interaction $\hat W(\r-\r')$.
The frequencies of the trap satisfy, without loss of generality, $\omega_x < \omega_y < \omega_z$.
We work with dimensionless quantities, $\hbar=m=1$.
Using Jacobi coordinates,
$\Q_k = \frac{1}{\sqrt{k(k+1)}} \sum_{j=1}^k (\r_{k+1} - \r_j), \ k=1,\ldots,N-1$,
$\Q_N = \frac{1}{\sqrt{N}} \sum_{j=1}^N \r_j$,
where $\r_1,\ldots,\r_N$ are the coordinates in the laboratory frame,
the Hamiltonian can be written as:
\beqn\label{HAM_3D}
\hat H(\Q_1,\ldots,\Q_N) = -\frac{1}{2} \frac{\partial^2}{\partial \Q^2_N} + 
\frac{1}{2} \left( \omega_x^2 Q^2_{N,x} + \omega_y^2 Q^2_{N,y} + \omega_z^2 Q^2_{N,z} \right) + 
\hat H_{rel}(\Q_1,\ldots,\Q_{N-1}).
\eeqn
The `relative' Hamiltonian $\hat H_{rel}$ collects all terms depending on the relative coordinates $\Q_1,\ldots,\Q_{N-1}$,
and $\Q_k=(Q_{k,x},Q_{k,y},Q_{k,z})$.
Suppose now that the bosons are prepared in the ground state of the non-interacting system.
The ground-state is separable in the Jacoby coordinates and reads
$\Phi(\Q_1,\ldots,\Q_N) = \left(\frac{\omega_x}{\pi}\right)^{\frac{N}{4}}\left(\frac{\omega_y}{\pi}\right)^{\frac{N}{4}}
\left(\frac{\omega_z}{\pi}\right)^{\frac{N}{4}}
e^{-\frac{1}{2}\sum_{k=1}^N \left(\omega_x Q_{k,x}^2 + \omega_y Q_{k,y}^2 + \omega_z Q_{k,z}^2\right)}$,
where the relation $\sum_{j=1}^N \r_j^2 = \sum_{k=1}^N \Q_k^2$ connecting the laboratory
and Jacoby coordinates is used.
The solution of the time-dependent many-boson Schr\"odinger equation,
$\hat H \Psi(t) = i \frac{\partial \Psi(t)}{\partial t}$,
where $\Phi$ is the initial condition,
reads $\Psi(t) = e^{-i \hat H t} \Phi$.
Consequently, 
because of the center-of-mass separability of the Hamiltonian $\hat H$
and of $\Phi$,
the position and momentum variances per particle
of the time-dependent state $\Psi(t)$,
for a general inter-particle interaction $\hat W(\r-\r')$,
are those of the static, non-interacting system:
\beqn\label{POS_MOM_VAR}
& &
{\rm Var|_{MB}}(\hat X) = \frac{1}{2\omega_x}, \quad
{\rm Var|_{MB}}(\hat Y) = \frac{1}{2\omega_y}, \quad
{\rm Var|_{MB}}(\hat Z) = \frac{1}{2\omega_z}, \quad \nonumber \\
& &
{\rm Var|_{MB}}(\hat P_X) = \frac{\omega_x}{2}, \quad
{\rm Var|_{MB}}(\hat P_Y) = \frac{\omega_y}{2}, \quad
{\rm Var|_{MB}}(\hat P_Z) = \frac{\omega_z}{2}. \quad \
\eeqn
In other words,
the anisotropies
${\rm Var|_{MB}}(\hat X) > {\rm Var|_{MB}}(\hat Y) > {\rm Var|_{MB}}(\hat Z)$
of the position operator
and
${\rm Var|_{MB}}(\hat P_Z) > {\rm Var|_{MB}}(\hat P_Y) > {\rm Var|_{MB}}(\hat P_X)$
of the momentum operator,
when computed at the many-body level of theory,
hold for all times during the out-of-equilibrium dynamics,
see the constant-value (dashed) curves in Figs.~\ref{f1} and \ref{f2}.

What happens at the Gross-Pitaevskii level of theory?
Can the mean-field variances have different orderings than the many-body variances,
i.e., belong to other anisotropy classes based on the $S_3$ permutation group,
see (\ref{VAR_ANI_3D_GP}),
than to $\{1\}$?
If yes, then why and how?
The Gross-Pitaevskii or non-linear Schr\"odinger equation
is given by\break\hfill
$\left[-\frac{1}{2}\frac{\partial^2}{\partial \r^2} +
\frac{1}{2} \left( \omega_x^2 x^2 + \omega_y^2 y^2 + \omega_z^2 z^2 \right)
+ g|\psi_{GP}(\r,t)|^2 \right] \psi_{GP}(\r,t) =
i\frac{\partial \psi_{GP}(\r,t)}{\partial t}$,
where $g=4\pi N a_s$ is the coupling constant and $a_s$
the $s$-wave scattering length of the
above two-body interaction $\hat W(\r-\r')$.
The initial condition, as above, is the ground state of the non-interacting system,
$\psi_{GP}(\r,0) = \left(\frac{\omega_x}{\pi}\right)^{\frac{1}{4}}\left(\frac{\omega_y}{\pi}\right)^{\frac{1}{4}}
\left(\frac{\omega_z}{\pi}\right)^{\frac{1}{4}}
e^{-\frac{1}{2}\left(\omega_x x^2 + \omega_y y^2 + \omega_z z^2\right)}$.
The Gross-Pitaevskii equation does not maintain the center-of-mass separability
of the initial condition because of its non-linear term,
which, therefore, can lead to variations of the position and momentum variances
when computed at the mean-field level of theory.

Figs.~\ref{f1} and \ref{f2} display the Gross-Pitaevskii dynamics of
the position and momentum variances per particle, respectively,
for four coupling constants,
$g=0.18, 9.0, 18.0$, and $27.0$.
To integrate the three-dimensional Gross-Pitaevskii equation we
use a box of size $[-10,10) \times [-10,10) \times [-10,10)$, a Fourier-discrete-variable-representation with
$128^3$ grid points and periodic boundary conditions,
and the numerical implementation embedded in \cite{GP_in_MCTDHB_Package}.
The dynamics are computed for the four
coupling constants $g$ and depicted 
by the oscillating (solid) curves in Figs.~\ref{f1} and \ref{f2}.
The left columns are for a 10\% anisotropy of the harmonic trap,
i.e., $\omega_x = 0.9$,
$\omega_y = 1.0$, and
$\omega_z = 1.1$,
and the right columns are for a 20\% anisotropy of the harmonic trap,
namely, $\omega_x = 0.8$,
$\omega_y = 1.0$, and
$\omega_z = 1.2$.
We remark that the expectation values per particle of the position ($\hat X, \hat Y, \hat Z$)
and momentum ($\hat P_X, \hat P_Y, \hat P_Z$) operators
computed at the mean-field and many-body levels of theory
coincide at the limit of an infinite number of particles
and are all equal to zero in the present scenario.

For the smallest coupling constant, $g=0.18$,
we see that the mean-field variances
oscillate with very small amplitudes around the respective constant values of the many-body variances.
This means that the mean-field anisotropy of the position variance,
${\rm Var|_{GP}}(\hat X) > {\rm Var|_{GP}}(\hat Y) > {\rm Var|_{GP}}(\hat Z)$,
and its many-body anisotropy,
${\rm Var|_{MB}}(\hat X) > {\rm Var|_{MB}}(\hat Y) > {\rm Var|_{MB}}(\hat Z)$,
are alike.
A similar situation is found for the momentum variance,
namely,
that the mean-field momentum anisotropy,
${\rm Var|_{GP}}(\hat P_Z) > {\rm Var|_{GP}}(\hat P_Y) > {\rm Var|_{GP}}(\hat P_X)$,
and the many-body anisotropy,
${\rm Var|_{MB}}(\hat P_Z) > {\rm Var|_{MB}}(\hat P_Y) > {\rm Var|_{MB}}(\hat P_X)$,
are the same.
Consequently,
we may conclude that
for small coupling constants the anisotropy class of the position operator is $\{1\}$
and, likewise, the anisotropy class of the momentum operator is $\{1\}$,
see (\ref{VAR_ANI_3D}).

The situation becomes more interesting for the larger coupling constants,
$g=9.0, 18.0$, and $27.0$.
We begin with the position variances, Fig.~\ref{f1}.
The variances are found to oscillate prominently,
with much larger amplitudes than for $g=0.18$,
and, subsequently, to cross each other.
There are three ingredients that enable and govern this crossing dynamics.
The first, is that the amplitudes of oscillations 
of ${\rm Var|_{GP}}(\hat X)$, ${\rm Var|_{GP}}(\hat Y)$, and ${\rm Var|_{GP}}(\hat Z)$ 
are slightly different already at short times,
with the former being the larger
and the latter being the smaller
(more prominent for 20\% than for 10\% trap anisotropy).
The second, is that the respective frequencies of oscillations
are also slightly different at short times,
with the former being the smaller
and the latter being the larger.
Both features correlate with the 
ordering of the frequencies of the trap,
$\omega_x < \omega_y < \omega_z$.
The third ingredient is that the three Cartesian components are coupled to
each other during the dynamics,
what impacts the oscillatory pattern at intermediate and later times
(more prominent for 10\% than for 20\% trap anisotropy, see Fig.~\ref{f1}).  

Combing the above,
we find for the 10\% trap anisotropy that
around $t=4.0$
${\rm Var|_{GP}}(\hat Z) > {\rm Var|_{GP}}(\hat Y) > {\rm Var|_{GP}}(\hat X)$ takes place,
around $t=6.0$
${\rm Var|_{GP}}(\hat X) > {\rm Var|_{GP}}(\hat Y) > {\rm Var|_{GP}}(\hat Z)$ holds (again),
and
around $t=7.0$
${\rm Var|_{GP}}(\hat Z) > {\rm Var|_{GP}}(\hat Y) > {\rm Var|_{GP}}(\hat X)$ occurs.
In other words,
the anisotropy class of the position variance starts as $\{1\}$ for $t=0$,
changes to $\{1,2\}$ around $t=4$,
is back to $\{1\}$ around $t=6$,
and becomes $\{1,2,3\}$ around $t=7$.
Furthermore,
this pattern is found to be robust for different, increasing coupling constants, see Fig.~\ref{f1}.
For 20\% trap anisotropy we find a different crossing patten of the position variances.
The anisotropy class begins as $\{1\}$ for $t=0$,
changes at around $t=3.0$ to $\{1,2\}$,
and immediately after,
at around $t=3.75$, it is $\{1,2,3\}$.
Now, around $t=7.0$ there is a broad regime 
of anisotropy class $\{1,2\}$.
Another difference of a geometrical origin between the
dynamics in the 20\% and 10\% trap anisotropies
can be seen for $g=9$, see Fig.~\ref{f1}c,d.
Here, the coupling constant is sufficiently large
to lead to crossing of all position variances for the 10\% anisotropy trap,
and, consequently,
to the position anisotropy class $\{1,2,3\}$
(at around $t=7$).
On the other hand,
for the 20\% anisotropy trap
the coupling constant is just short
of allowing all position variances to cross each other
and, clearly, the anisotropy class $\{1,2,3\}$ cannot occur
(as it happens at around $t=3.75$ for the further larger coupling constants, $g=18$ and $27$).
All in all,
we have demonstrated in a rather common (out-of-equilibrium quench) scenario
the emergence of anisotropy classes other than $\{1\}$,
i.e., $\{1,2\}$ and $\{1,2,3\}$,
for the position operator
of a Bose-Einstein condensate
at the infinite-particle-number limit.

The results for the momentum variances per particle, see Fig.~\ref{f2},
follow similar and corresponding trends as those for the position operator,
albeit the crossings of the respective momentum curves
take place during slightly narrower time windows than for the position operator
for the parameters used.
Thus, for the 10\% trap anisotropy we have
${\rm Var|_{GP}}(\hat P_X) > {\rm Var|_{GP}}(\hat P_Y) > {\rm Var|_{GP}}(\hat P_Z)$ around $t=4.25$,
${\rm Var|_{GP}}(\hat P_Z) > {\rm Var|_{GP}}(\hat P_Y) > {\rm Var|_{GP}}(\hat P_X)$ in the vicinity of $t=6.0$,
and
${\rm Var|_{GP}}(\hat P_Y) > {\rm Var|_{GP}}(\hat P_X) > {\rm Var|_{GP}}(\hat P_Z)$ around $t=7.25$.
Therefore,
the anisotropy class of the momentum variance starts at $\{1\}$ for $t=0$,
turns to $\{1,2\}$ around $t=4.25$,
returns to $\{1\}$ for a wider time window around $t=6.0$,
and changes to $\{1,2,3\}$ around $t=7.25$.
For the 20\% trap anisotropy we find,
starting from the anisotropy class $\{1\}$ for $t=0$,
the class $\{1,2,3\}$ at around $t=4.0$,
the class $\{1,2\}$ at around $t=4.75$,
again the class $\{1\}$ at around $t=6$,
and once more the class $\{1,2,3\}$ at around $t=9.5$.
As for the position variance,
we find the pattern to be robust for different, increasing coupling constants, see Fig.~\ref{f2}.
Furthermore,
the above-discussed difference of a geometrical origin between the
position-variance dynamics in the 20\% and 10\% trap anisotropies for $g=9$
emerges also for the momentum variance, see Fig.~\ref{f2}c,d.
Here, the coupling constant is sufficiently large
to lead to crossing of all momentum variances for the 10\% trap anisotropy,
but not for the 20\% trap anisotropy.
As a result,
the former system exhibits also the anisotropy class $\{1,2,3\}$ for the momentum variance,
whereas the latter only the $\{1,2\}$ anisotropy class.
Summarizing,
we have demonstrated in a simple scenario, of an out-of-equilibrium breathing dynamics,
the emergence of anisotropy classes other than $\{1\}$,
namely, $\{1,2\}$ and $\{1,2,3\}$,
for the many-particle position as well as many-particle momentum operators
of a trapped Bose-Einstein condensate condensate at the limit of an infinite number of particles.

\subsection{Angular-momentum variance in the ground state of a
three-dimensional trapped Bose-Einstein condensate}\label{angular}

The possibility to learn on the relations governing correlations
and variance anisotropy between the different components
of the angular-momentum operator opens up only in three spatial dimensions.
Here, in the context of the present work, the challenge is to find a many-particle model
where angular-momentum properties can be treated analytically
at the many-body level of theory and
in the limit of an infinite number of particles.
Such a model is the three-dimensional anisotropic harmonic-interaction model,
and the results presented below build on and clearly extends
the investigation of the two-dimensional anisotropic harmonic-interaction model
reported in \cite{VAR6}.
The harmonic-interaction model has been used quite extensively including to model Bose-Einstein condensates
\cite{H0,H1,H2,H3,H4,H5,H6,H7,H8,H9,H10,H11,H12,H13,H14}.
Finally, and as a bonus, we mention that the 
three-dimensional anisotropic harmonic-interaction model
can also be solved analytically at the mean-field level of theory, which is useful for the analysis.

In the laboratory frame
the three-dimensional anisotropic harmonic-interaction model reads:
$\hat H(\r_1,\ldots,\r_N) = 
\sum_{j=1}^N \left[-\frac{1}{2}\frac{\partial^2}{\partial \r_j^2} + \frac{1}{2}
\left(\omega_x^2 x_j^2 + \omega_y^2 y_j^2 + \omega_z^2 z_j^2 \right)\right] 
+ \lambda_0 \sum_{1 \le j < k}^N (\r_j-\r_k)^2$, i.e.,
it is obtained from the Hamiltonian (\ref{HAM_3D}) when the two-body interaction is $\hat W(\r-\r') = \lambda_0 (\r-\r')^2$.
Then, the `relative' Hamiltonian is given explicitly by
\beqn\label{HAM_3D_HIM}
& & \hat H_{rel}(\Q_1,\ldots,\Q_{N-1}) = \\
& & \quad = \sum_{k=1}^{N-1} \left[
-\frac{1}{2} \frac{\partial^2}{\partial \Q^2_k} + 
\frac{1}{2} \left( \sqrt{\omega_x^2+2N\lambda_0} Q^2_{k,x} + \sqrt{\omega_y^2+2N\lambda_0} Q^2_{k,y} + \sqrt{\omega_z^2+2N\lambda_0} Q^2_{k,z} \right) \right]. \nonumber \
\eeqn
The many-body ground state of $\hat H$ is readily obtained and given by
\beqn\label{HIM_3D_GS}
& & \Psi(\Q_1,\ldots,\Q_N) = \nonumber \\
& & = \left(\frac{\omega_x}{\pi}\right)^{\frac{1}{4}}
\left(\frac{\omega_y}{\pi}\right)^{\frac{1}{4}}
\left(\frac{\omega_z}{\pi}\right)^{\frac{1}{4}}
\left(\frac{\sqrt{\omega_x^2+2N\lambda_0}}{\pi}\right)^{\frac{N-1}{4}}
\left(\frac{\sqrt{\omega_y^2+2N\lambda_0}}{\pi}\right)^{\frac{N-1}{4}} 
\left(\frac{\sqrt{\omega_z^2+2N\lambda_0}}{\pi}\right)^{\frac{N-1}{4}} 
\times \nonumber \\
& &
e^{-\frac{1}{2} \left(\sqrt{\omega_x^2+2N\lambda_0} \sum_{k=1}^{N-1} Q_{k,x}^2 + \omega_x Q_{N,x}^2 \right)}
e^{-\frac{1}{2} \left( \sqrt{\omega_y^2+2N\lambda_0} \sum_{k=1}^{N-1} Q_{k,y}^2 + \omega_y Q_{N,y}^2 \right)} 
e^{-\frac{1}{2} \left( \sqrt{\omega_z^2+2N\lambda_0} \sum_{k=1}^{N-1} Q_{k,z}^2 + \omega_z Q_{N,z}^2 \right)}.
\nonumber \\ \
\eeqn
As states above,
it is also possible to solve analytically
the three-dimensional anisotropic harmonic-interaction model
at the mean-field level of theory by generalizing \cite{VAR6,H3}.
The final result for
the mean-field solution of the ground state reads
\beqn\label{HIM_3D_GP}
& & \psi_{GP}(\r) =
\left(\frac{\sqrt{\omega_x^2+2\Lambda}}{\pi}\right)^{\frac{1}{4}}
\left(\frac{\sqrt{\omega_y^2+2\Lambda}}{\pi}\right)^{\frac{1}{4}} 
\left(\frac{\sqrt{\omega_z^2+2\Lambda}}{\pi}\right)^{\frac{1}{4}} \times \nonumber \\
& & \times
e^{-\frac{1}{2}\sqrt{\omega_x^2+2\Lambda} \, x^2}
e^{-\frac{1}{2}\sqrt{\omega_y^2+2\Lambda} \, y^2}
e^{-\frac{1}{2}\sqrt{\omega_z^2+2\Lambda} \, z^2},
\eeqn
where $\Lambda=\lambda_0(N-1)$ is the interaction parameter.
For reference,
$\psi_{GP}(\r)$ solves the Gross-Pitaevskii equation
$\left[-\frac{1}{2}\frac{\partial^2}{\partial \r^2} +
\frac{1}{2} \left( \omega_x^2 x^2 + \omega_y^2 y^2 + \omega_z^2 z^2 \right) +
\Lambda \int d\r' |\psi_{GP}(\r')|^2 (\r-\r')^2 \right] \psi_{GP}(\r) =
\mu \psi_{GP}(\r)$,
where $\mu$ is the chemical potential.
Note that both many-body and mean-field solutions can be written as products of
the respective solutions in one dimension along the $x$, $y$, and $z$ directions.

Before we arrive at the angular-momentum variances and for our needs, see below,
we make a stopover and compute the position and momentum variances per particle in the model.
At the many-body level we obviously have the result (\ref{POS_MOM_VAR}),
since for the interacting ground-state the center-of-mass is separable and, hence,
the position and momentum variances are independent of the two-body interaction.
At the mean-field level we readily find from (\ref{HIM_3D_GP}) the result
\beqn\label{POS_MOM_VAR_GP}
& &
{\rm Var|_{GP}}(\hat X) = \frac{1}{2\sqrt{\omega_x^2+2\Lambda}}, \quad
{\rm Var|_{GP}}(\hat Y) = \frac{1}{2\sqrt{\omega_y^2+2\Lambda}}, \quad
{\rm Var|_{GP}}(\hat Z) = \frac{1}{2\sqrt{\omega_z^2+2\Lambda}}, \quad \nonumber \\
& &
{\rm Var|_{GP}}(\hat P_X) = \frac{\sqrt{\omega_x^2+2\Lambda}}{2}, \quad
{\rm Var|_{GP}}(\hat P_Y) = \frac{\sqrt{\omega_y^2+2\Lambda}}{2}, \quad
{\rm Var|_{GP}}(\hat P_Z) = \frac{\sqrt{\omega_z^2+2\Lambda}}{2}. \quad \
\eeqn
The mean-field variances (\ref{POS_MOM_VAR_GP}) 
depend on the interaction parameter $\Lambda$,
unlike the respective many-body variances (\ref{POS_MOM_VAR}).
It turns out that this property would be instrumental when analyzing
the anisotropy of the angular-momentum variance below.
We briefly comment on the anisotropies of the position and momentum variances
in the model.
Comparing the mean-field (\ref{POS_MOM_VAR_GP}) and
many-body (\ref{POS_MOM_VAR}) variances per particle
we find that the former belong to the anisotropy class $\{1\}$
independently of the interaction parameter $\Lambda$
both for the position and momentum operators.
For the mean-field variance of the ground state at the infinite-particle-number limit
to belong to an anisotropy class other than $\{1\}$,
one would have to go beyond the simple single-well geometry,
see the anisotropy of the position variance in a double-well potential
in two spatial dimensions \cite{VAR7}.

We can now move to the expressions for the angular-momentum variances
at the limit of an infinite number of particles,
by generalizing results obtained in two spatial dimensions \cite{VAR6} two three spatial dimensions.
The calculation at the mean-field level using (\ref{HIM_3D_GP}) readily gives
\beqn\label{VAR_Lx_Ly_Lz_GP}
& &
{\rm Var|_{GP}}(\hat L_X) = \frac{1}{4} \frac{\left(\sqrt{\omega_y^2+2\Lambda}-\sqrt{\omega_z^2+2\Lambda}\right)^2}{\sqrt{\omega_y^2+2\Lambda} \sqrt{\omega_z^2+2\Lambda}}
= \frac{1}{4}\frac{\left(\frac{\sqrt{\omega_y^2+2\Lambda}}{\sqrt{\omega_z^2+2\Lambda}}-1\right)^2}{\frac{\sqrt{\omega_y^2+2\Lambda}}{\sqrt{\omega_z^2+2\Lambda}}}, \nonumber \\
& &
{\rm Var|_{GP}}(\hat L_Y) = \frac{1}{4} \frac{\left(\sqrt{\omega_z^2+2\Lambda}-\sqrt{\omega_x^2+2\Lambda}\right)^2}{\sqrt{\omega_z^2+2\Lambda} \sqrt{\omega_x^2+2\Lambda}}
= \frac{1}{4}\frac{\left(\frac{\sqrt{\omega_z^2+2\Lambda}}{\sqrt{\omega_x^2+2\Lambda}}-1\right)^2}{\frac{\sqrt{\omega_z^2+2\Lambda}}{\sqrt{\omega_x^2+2\Lambda}}}, \nonumber \\
& &
{\rm Var|_{GP}}(\hat L_Z) = \frac{1}{4} \frac{\left(\sqrt{\omega_x^2+2\Lambda}-\sqrt{\omega_y^2+2\Lambda}\right)^2}{\sqrt{\omega_x^2+2\Lambda} \sqrt{\omega_y^2+2\Lambda}}
= \frac{1}{4}\frac{\left(\frac{\sqrt{\omega_x^2+2\Lambda}}{\sqrt{\omega_y^2+2\Lambda}}-1\right)^2}{\frac{\sqrt{\omega_x^2+2\Lambda}}{\sqrt{\omega_y^2+2\Lambda}}}. \
\eeqn
In the absence of interaction these
expressions boil down, respectively, to
$\frac{1}{4}\frac{\left(\frac{\omega_y}{\omega_z}-1\right)^2}{\frac{\omega_y}{\omega_z}}$,
$\frac{1}{4}\frac{\left(\frac{\omega_z}{\omega_x}-1\right)^2}{\frac{\omega_z}{\omega_x}}$,
and
$\frac{1}{4}\frac{\left(\frac{\omega_x}{\omega_y}-1\right)^2}{\frac{\omega_x}{\omega_y}}$,
the angular-momentum variances of a single particle in a three-dimensional anisotropic harmonic potential.
We see that for non-interacting particles and at the mean-field level
the angular-momentum variances per particle depend on the ratios of frequencies,
not on their absolute magnitudes.
In the first case these are the bare frequencies of the harmonic trap
whereas in the second case these are the interaction-dressed frequencies (\ref{HIM_3D_GP})
resulting from the non-linear term.

The computation of the many-body variances is lengthier and using \cite{VAR6} the final expressions for
the correlations terms (\ref{VAR_INF}) are
\beqn\label{VAR_Lx_Ly_Lz_CORR}
& & {\rm Var|_{correlations}}(\hat L_X) =
\frac{1}{4} \frac{\left(\sqrt{\omega_y^2+2\Lambda}-\sqrt{\omega_z^2+2\Lambda}\right)^2}{\sqrt{\omega_y^2+2\Lambda} \sqrt{\omega_z^2+2\Lambda}}
\left(\sqrt{1+\frac{2\Lambda}{\omega_y^2}} - 1\right)\left(\sqrt{1+\frac{2\Lambda}{\omega_z^2}} - 1\right), \nonumber \\
& & {\rm Var|_{correlations}}(\hat L_Y) =
\frac{1}{4} \frac{\left(\sqrt{\omega_z^2+2\Lambda}-\sqrt{\omega_x^2+2\Lambda}\right)^2}{\sqrt{\omega_z^2+2\Lambda} \sqrt{\omega_x^2+2\Lambda}}
\left(\sqrt{1+\frac{2\Lambda}{\omega_z^2}} - 1\right)\left(\sqrt{1+\frac{2\Lambda}{\omega_x^2}} - 1\right), \nonumber \\
& & {\rm Var|_{correlations}}(\hat L_Z) =
\frac{1}{4} \frac{\left(\sqrt{\omega_x^2+2\Lambda}-\sqrt{\omega_y^2+2\Lambda}\right)^2}{\sqrt{\omega_x^2+2\Lambda} \sqrt{\omega_y^2+2\Lambda}}
\left(\sqrt{1+\frac{2\Lambda}{\omega_x^2}} - 1\right)\left(\sqrt{1+\frac{2\Lambda}{\omega_y^2}} - 1\right). \nonumber \\
\eeqn
Hence, adding (\ref{VAR_Lx_Ly_Lz_GP}) and (\ref{VAR_Lx_Ly_Lz_CORR}) we readily have from (\ref{VAR_INF})
the many-body variances per particle at the infinite-particle-number limit,
${\rm Var|_{MB}}(\hat L_X)$,
${\rm Var|_{MB}}(\hat L_Y)$,
and
${\rm Var|_{MB}}(\hat L_Z)$.
We remark that the expectation values per particle of the angular-momentum operator ($\hat L_X, \hat L_Y, \hat L_Z$),
as well as the respective expectation values of the position and momentum operators,
computed at the mean-field and many-body levels of theory
coincide at the limit of an infinite number of particles
and are all equal to zero in the ground state.

We investigate and discuss an example.
Let the frequencies of the three-dimensional anisotropic harmonic trap be
$\omega_x = 0.7$,
$\omega_y = 5.0$,
and
$\omega_z = 10.5$.
Their ratios from large to small are:
$\frac{\omega_z}{\omega_x} = 15$,
$\frac{\omega_y}{\omega_x} = 7\frac{1}{7}$,
and
$\frac{\omega_z}{\omega_y} = 2\frac{1}{10}$.
Then,
the values of the angular-momentum variances per particle
at zero interaction parameter, $\Lambda=0$, are given from large to small by
${\rm Var|_{GP}}(\hat L_Y) = {\rm Var|_{MB}}(\hat L_Y) = \frac{7^2}{15} \approx 3.267$,
${\rm Var|_{GP}}(\hat L_Z) = {\rm Var|_{MB}}(\hat L_Z) = \frac{43^2}{1400} \approx 1.321$,
and
${\rm Var|_{GP}}(\hat L_X) = {\rm Var|_{MB}}(\hat L_X) = \frac{11^2}{840} \approx 0.144$.
Indeed, as the ratio of frequencies with respect to two axes is bigger,
the corresponding angular-momentum variance per particle with respect to the third axis is larger,
and vise versa.

\begin{figure}[!]
\begin{center}
\hglue -1.0 truecm
\includegraphics[width=0.345\columnwidth,angle=-90]{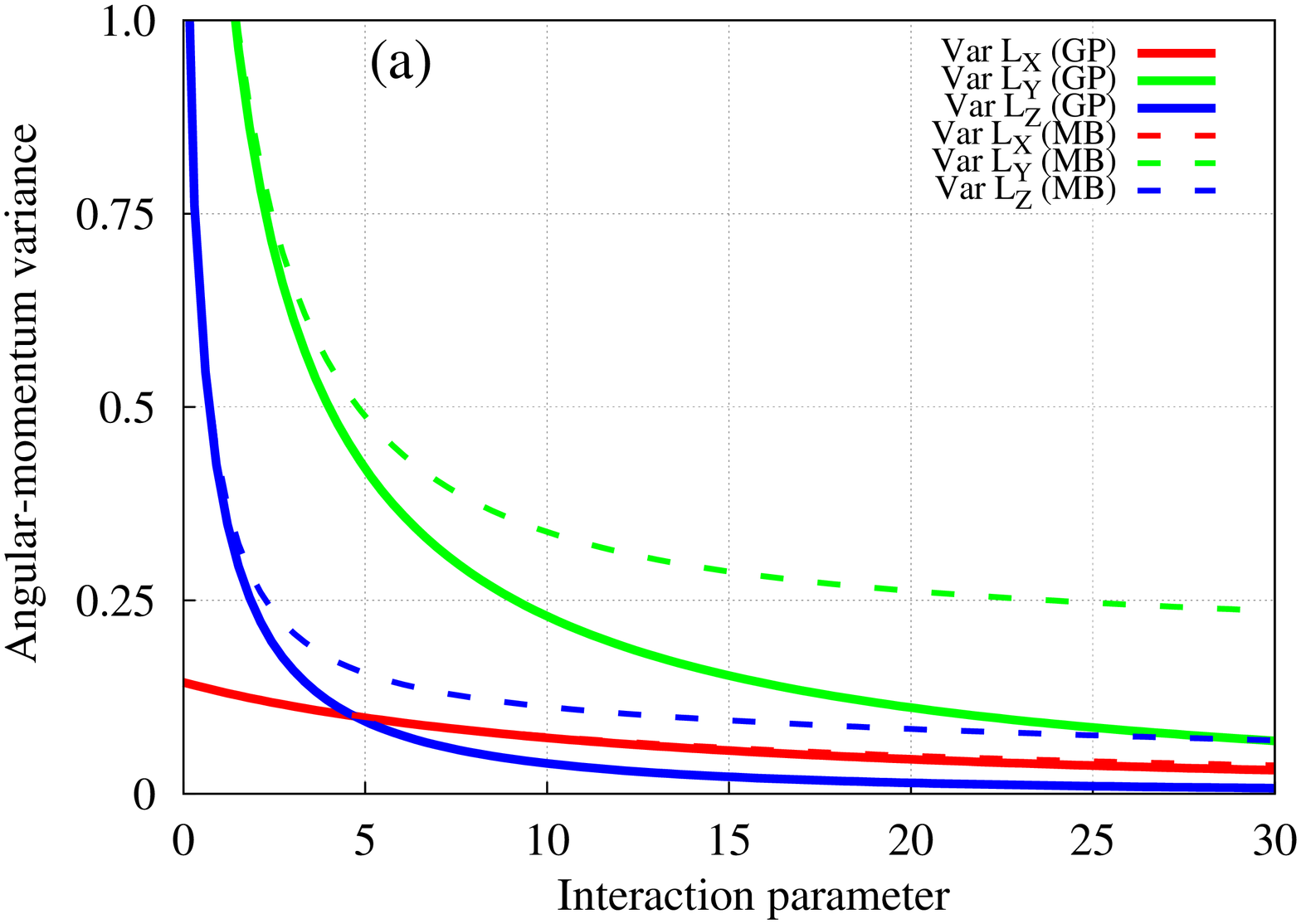}
\includegraphics[width=0.345\columnwidth,angle=-90]{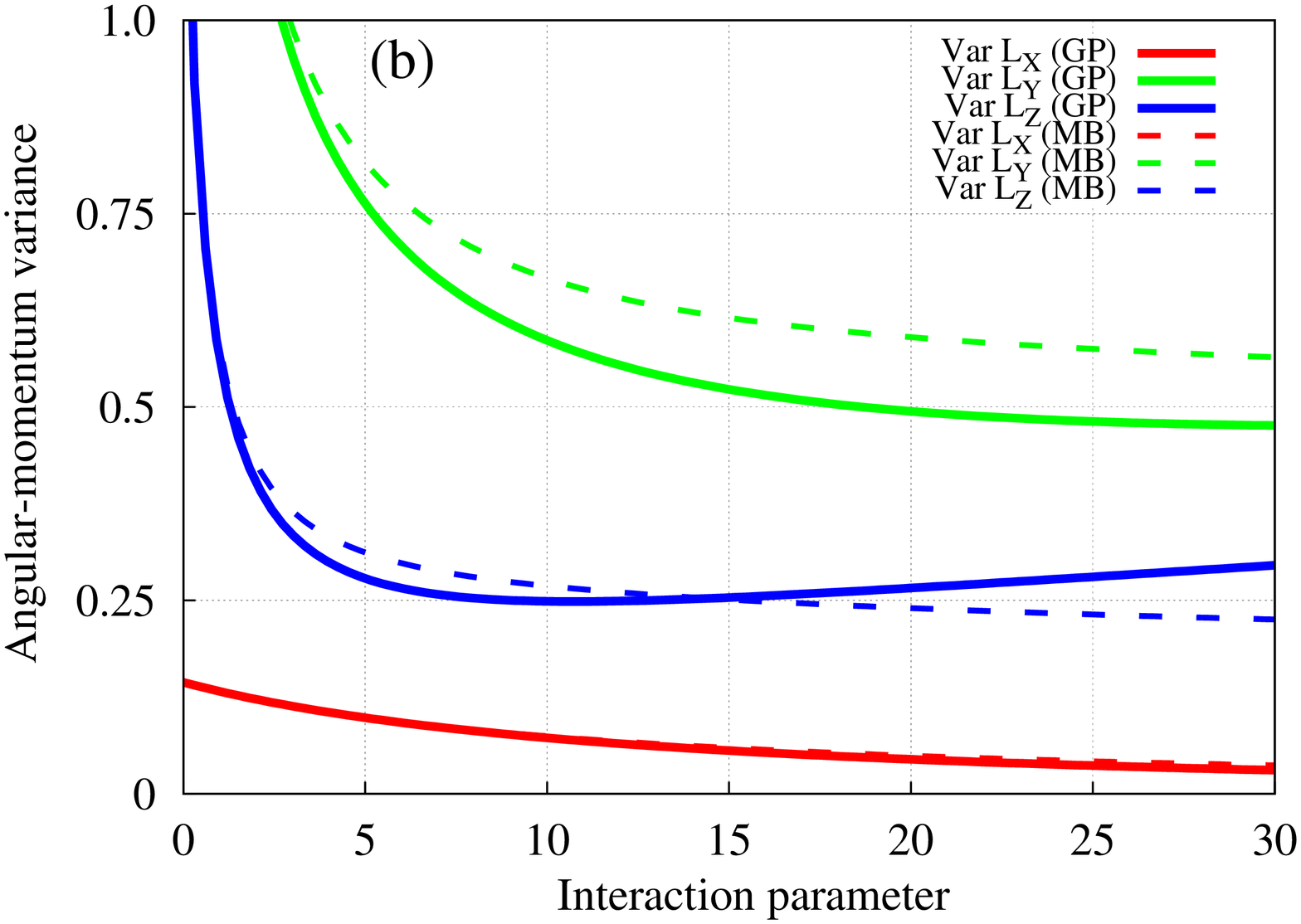}
\hglue -1.0 truecm
\includegraphics[width=0.345\columnwidth,angle=-90]{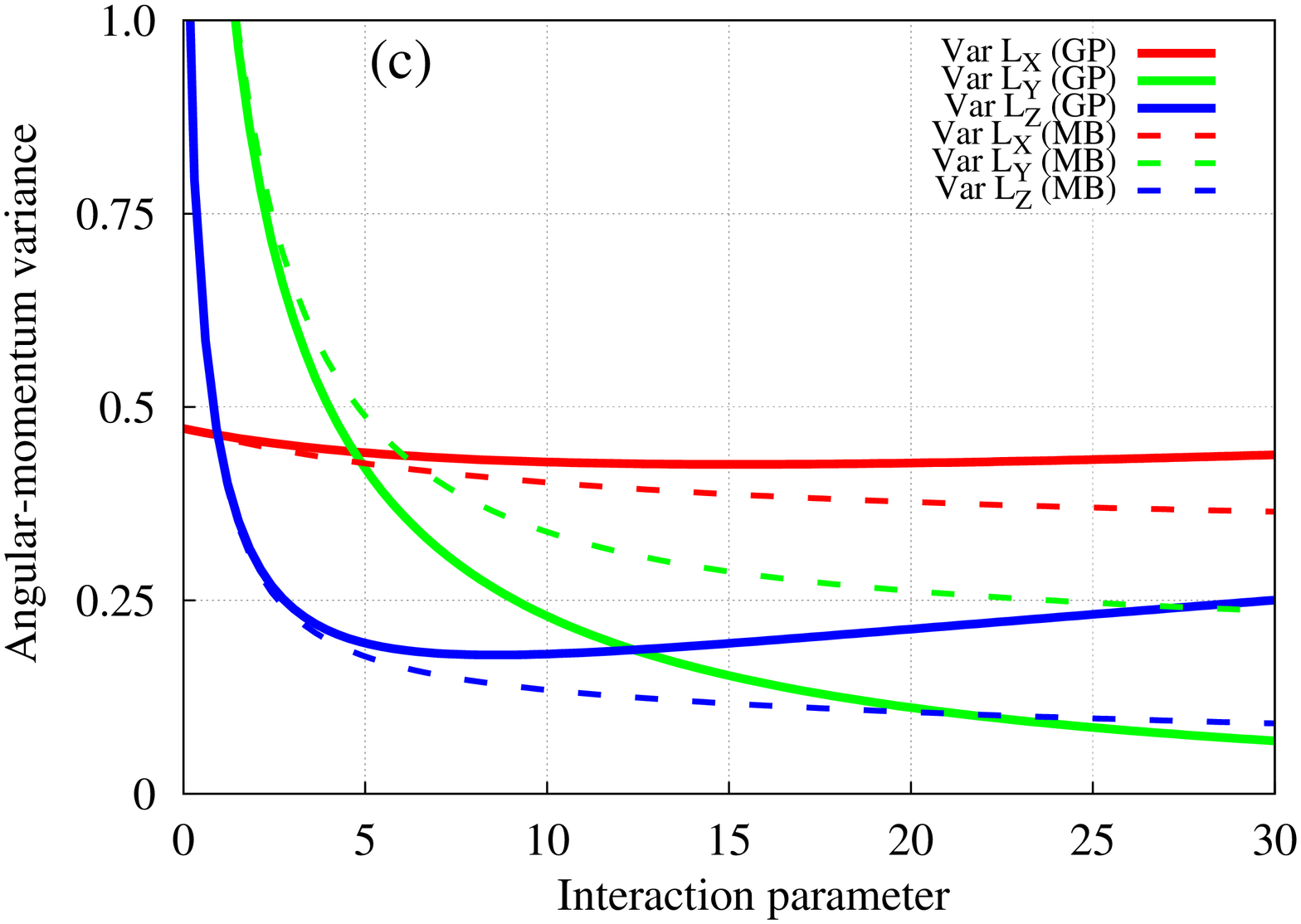}
\includegraphics[width=0.345\columnwidth,angle=-90]{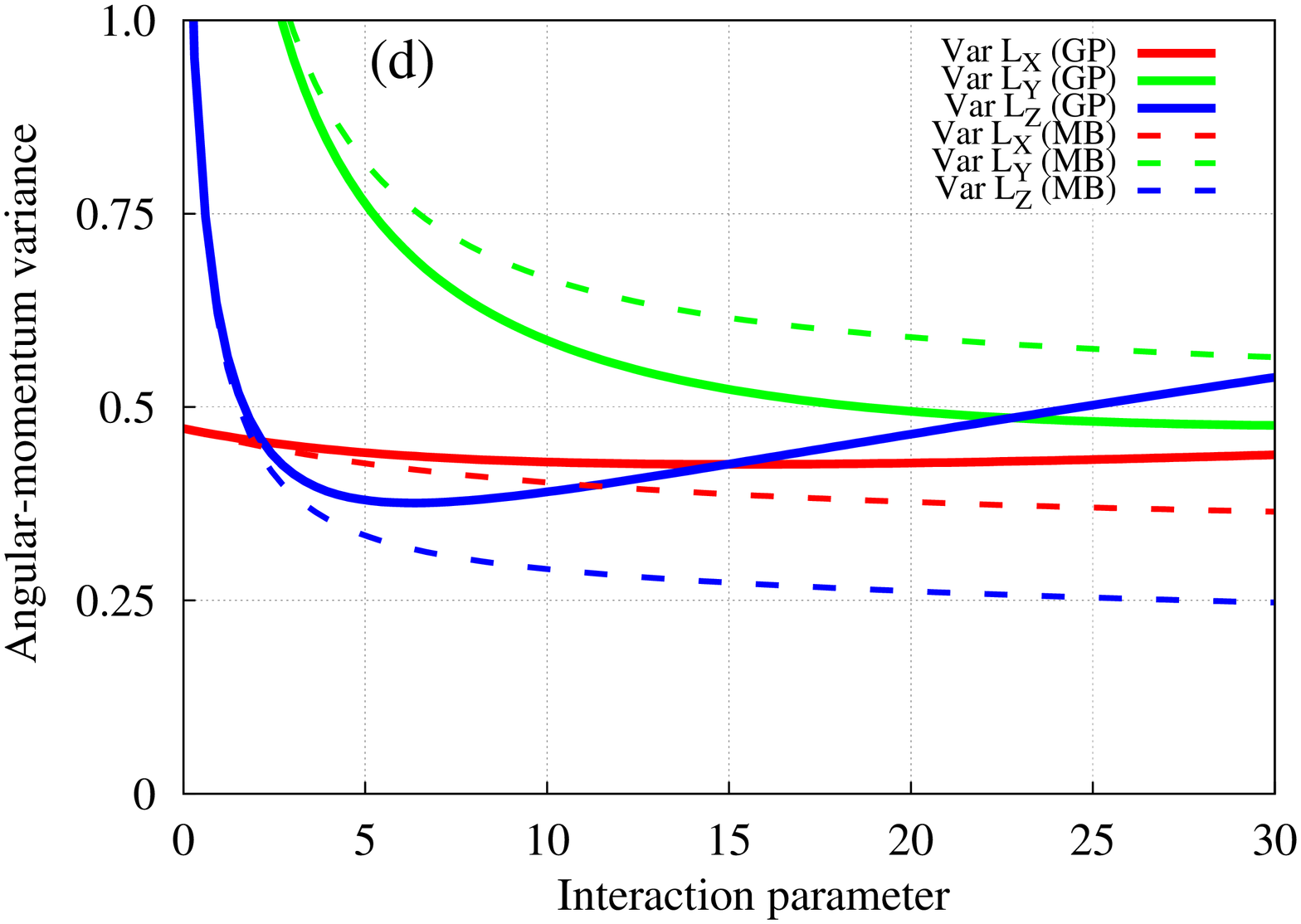}
\end{center}
\vglue 0.75 truecm
\caption{Many-particle angular-momentum ($\hat L_X$, $\hat L_Y$, and $\hat L_Z$; in red, green, and blue)
variance per particle as a function of the interaction parameter $\Lambda$
computed at the limit of an infinite number of particles
within many-body (dashed lines) and mean-field (solid lines) levels of theory for the ground state
of the three-dimensional anisotropic harmonic-interaction model.
The frequencies of the trap are $\omega_x = 0.7$, $\omega_y = 5.0$, and $\omega_z = 10.5$.
Results at several translations $\r_0$ of the center of the trap are shown in the panels:
(a) $\r_0=(0,0,0)$;
(b) $\r_0=(0.25,0,0)$;
(c) $\r_0=(0,0.25,0)$;
(d) $\r_0=(0.25,0.25,0)$.
Different anisotropy classes of the angular-momentum variance emerge with the interaction parameter.
See the text for more details.
The quantities shown are dimensionless.}
\label{f3}
\end{figure}

What happens as the interaction sets in?
Fig.~\ref{f3}a depicts the many-body and mean-field
angular-momentum variances as a function of the interaction parameter parameter $\Lambda$.
We examine positive values of $\Lambda$ which correspond to the attractive sector of the 
harmonic-interaction model, see, e.g., \cite{VAR6,H3,H7}.
Let us analyze the observations.
With increasing interaction parameter the density narrows, along the $x$, $y$, and $z$ directions.
This is clear because the interaction between particles is attractive,
and is manifested by the monotonously decreasing values of the 
position variances per particle (\ref{POS_MOM_VAR_GP}).
Furthermore,
the density becomes less anisotropic,
because the ratios of the dressed frequencies
$\frac{\sqrt{\omega_z^2+2\Lambda}}{\sqrt{\omega_x^2+2\Lambda}}$,
$\frac{\sqrt{\omega_y^2+2\Lambda}}{\sqrt{\omega_x^2+2\Lambda}}$,
and
$\frac{\sqrt{\omega_z^2+2\Lambda}}{\sqrt{\omega_y^2+2\Lambda}}$
monotonously decreases with increasing $\Lambda$.
Consequently, the angular-momentum variances per particle decrease with the interaction parameter as well,
see Fig.~\ref{f3}a.
The mean-field angular-momentum variances (\ref{VAR_Lx_Ly_Lz_GP}) are monotonously decreasing
because of the just-described decreasing ratios of the dressed frequencies.
The many-body angular-momentum variances are decreasing,
at least for the values of interaction parameters studied here,
because the positive-value correlations terms (\ref{VAR_Lx_Ly_Lz_CORR})
grow slower 
than the mean-field angular-momentum variances decrease with $\Lambda$.

All in all, the anisotropy of the angular-momentum variance can now be determined.
We find the anisotropy 
${\rm Var|_{MB}}(\hat L_Y) > {\rm Var|_{MB}}(\hat L_Z) > {\rm Var|_{MB}}(\hat L_X)$
to hold for all interaction parameters at the many-body level of theory.
At the mean-field level of theory we find the same anisotropy, namely,
${\rm Var|_{GP}}(\hat L_Y) > {\rm Var|_{GP}}(\hat L_Z) > {\rm Var|_{GP}}(\hat L_X)$,
to hold for small interaction parameters.
But then, at just about $\Lambda=5.0$ the mean-field anisotropy changes to
${\rm Var|_{GP}}(\hat L_Y) > {\rm Var|_{GP}}(\hat L_X) > {\rm Var|_{GP}}(\hat L_Z)$,
and this anisotropy continues for larger interaction parameters. 
Hence, we have found that 
the anisotropy of the angular momentum operator
in the ground state of the
three-dimensional anisotropic harmonic-interaction model
at the infinite-particle-number limit
changes as a function of the interaction parameter
from the anisotropy class $\{1\}$ to $\{1,2\}$,
see Fig.~\ref{f3}a.

Still, can the above-found picture of angular-momentum variance anisotropy be made richer?
The answer is positive and requires one to dive deeper into to properties of
angular-momentum variances under translations.
To this end,
we employ and extend prior work in two spatial dimensions \cite{VAR6}.
Suppose now that the harmonic trap is located not in the origin but translated to a general point $\r_0=(x_0,y_0,z_0)$.
The expectation values per particle of the momentum and angular-momentum operators are still zero,
whereas the expectation value of the position operator is, of course, $\r_0$.
Whereas the position and momentum variances are invariant to translations,
the angular-momentum variances are not, which open up another degree of freedom to investigate the anisotropy class
of the angular-momentum variances in three spatial dimensions. 
Using the transformation properties of the angular-momentum operator $(\hat L_X, \hat L_Y, \hat L_Z)$ under translations,
see appendix \ref{APP} and \cite{VAR6},
the final expressions
for the translated angular-momentum variances per particle
read explicitly
\beqn\label{VAR_Lx_Ly_Lz_MB_TRANS_FINAL}
& &
{\rm Var|_{MB}}(\hat L_X;\r_0) = 
\frac{1}{4} \frac{\left(\sqrt{\omega_y^2+2\Lambda}-\sqrt{\omega_z^2+2\Lambda}\right)^2}{\sqrt{\omega_y^2+2\Lambda} \sqrt{\omega_z^2+2\Lambda}}
\left[1 +
\left(\sqrt{1+\frac{2\Lambda}{\omega_y^2}} - 1\right)\left(\sqrt{1+\frac{2\Lambda}{\omega_z^2}} - 1\right)\right] + 
\nonumber \\
& & \quad + y_0^2 \frac{\omega_z}{2} + z_0^2 \frac{\omega_y}{2},
\nonumber \\
& &
{\rm Var|_{MB}}(\hat L_Y;\r_0) = 
\frac{1}{4} \frac{\left(\sqrt{\omega_z^2+2\Lambda}-\sqrt{\omega_x^2+2\Lambda}\right)^2}{\sqrt{\omega_z^2+2\Lambda} \sqrt{\omega_x^2+2\Lambda}}
\left[1 +
\left(\sqrt{1+\frac{2\Lambda}{\omega_z^2}} - 1\right)\left(\sqrt{1+\frac{2\Lambda}{\omega_x^2}} - 1\right)\right] +
\nonumber \\
& & \quad + z_0^2 \frac{\omega_x}{2} + x_0^2 \frac{\omega_z}{2},
\nonumber \\
& &
{\rm Var|_{MB}}(\hat L_Z;\r_0) = 
\frac{1}{4} \frac{\left(\sqrt{\omega_x^2+2\Lambda}-\sqrt{\omega_y^2+2\Lambda}\right)^2}{\sqrt{\omega_x^2+2\Lambda} \sqrt{\omega_y^2+2\Lambda}}
\left[1 +
\left(\sqrt{1+\frac{2\Lambda}{\omega_x^2}} - 1\right)\left(\sqrt{1+\frac{2\Lambda}{\omega_y^2}} - 1\right)\right] +
\nonumber \\
& & \quad + x_0^2 \frac{\omega_y}{2} + y_0^2 \frac{\omega_x}{2}
\eeqn
at the many-body level of theory and
\beqn\label{VAR_Lx_Ly_Lz_GP_TRANS_FINAL}
& &
{\rm Var|_{GP}}(\hat L_X;\r_0) = \frac{1}{4} \frac{\left(\sqrt{\omega_y^2+2\Lambda}-\sqrt{\omega_z^2+2\Lambda}\right)^2}{\sqrt{\omega_y^2+2\Lambda} \sqrt{\omega_z^2+2\Lambda}} +
y_0^2 \frac{\sqrt{\omega_z^2+2\Lambda}}{2} + z_0^2 \frac{\sqrt{\omega_y^2+2\Lambda}}{2}, \nonumber \\
& &
{\rm Var|_{GP}}(\hat L_Y;\r_0) = \frac{1}{4} \frac{\left(\sqrt{\omega_z^2+2\Lambda}-\sqrt{\omega_x^2+2\Lambda}\right)^2}{\sqrt{\omega_z^2+2\Lambda} \sqrt{\omega_x^2+2\Lambda}} +
z_0^2 \frac{\sqrt{\omega_x^2+2\Lambda}}{2} + x_0^2 \frac{\sqrt{\omega_z^2+2\Lambda}}{2}, \nonumber \\
& &
{\rm Var|_{GP}}(\hat L_Z;\r_0) = \frac{1}{4} \frac{\left(\sqrt{\omega_x^2+2\Lambda}-\sqrt{\omega_y^2+2\Lambda}\right)^2}{\sqrt{\omega_x^2+2\Lambda} \sqrt{\omega_y^2+2\Lambda}} +
x_0^2 \frac{\sqrt{\omega_y^2+2\Lambda}}{2} + y_0^2 \frac{\sqrt{\omega_x^2+2\Lambda}}{2} \
\eeqn
at the mean-field level of theory.
Let us examine expressions (\ref{VAR_Lx_Ly_Lz_MB_TRANS_FINAL}) and (\ref{VAR_Lx_Ly_Lz_GP_TRANS_FINAL}) 
more closely.
The terms added to the translated angular-momentum variances at the many-body level
depend on the corresponding components of the translation vector but not on the interaction parameter,
whereas the added terms at the mean-field level of theory depend on 
and increase with $\Lambda$, see appendix \ref{APP} for more details.
The combined effects can be seen in Fig.~\ref{f3},
compare panels Fig.~\ref{f3}b,c,d with panel Fig.~\ref{f3}a.

For the translation by $\r_0=(0.25,0,0)$,
the angular-momentum variances ${\rm Var|_{MB}}(\hat L_X;\r_0)$ and ${\rm Var|_{GP}}(\hat L_X;\r_0)$ 
are invariant quantities,
${\rm Var|_{MB}}(\hat L_Y;\r_0)$ and ${\rm Var|_{MB}}(\hat L_Z;\r_0)$
are shifted by interaction-independent values,
and
${\rm Var|_{GP}}(\hat L_Y;\r_0)$ and ${\rm Var|_{GP}}(\hat L_Z;\r_0)$
increase by interaction-dependent values,
see (\ref{VAR_Lx_Ly_Lz_MB_TRANS_FINAL}) and (\ref{VAR_Lx_Ly_Lz_GP_TRANS_FINAL}), 
and compare Fig.~\ref{f3}b and Fig.~\ref{f3}a.
The combined effect is that now
both
${\rm Var|_{MB}}(\hat L_Y;\r_0) > {\rm Var|_{MB}}(\hat L_Z;\r_0) > {\rm Var|_{MB}}(\hat L_X;\r_0)$
and
${\rm Var|_{GP}}(\hat L_Y;\r_0) > {\rm Var|_{GP}}(\hat L_Z;\r_0) > {\rm Var|_{GP}}(\hat L_X;\r_0)$
hold
for all interaction parameters $\Lambda$ in the range studied.
Consequently,
the anisotropy class of the angular-momentum variance per particle for $\r_0=(0.25,0,0)$
is $\{1\}$ only.

Next, 
for $\r_0=(0,0.25,0)$,
the angular-momentum variances ${\rm Var|_{MB}}(\hat L_Y;\r_0)$ and\break\hfill
${\rm Var|_{GP}}(\hat L_Y;\r_0)$ 
are invariant quantities,
${\rm Var|_{MB}}(\hat L_X;\r_0)$ and ${\rm Var|_{MB}}(\hat L_Z;\r_0)$
are shifted by interaction-independent values,
and
${\rm Var|_{GP}}(\hat L_X;\r_0)$ and ${\rm Var|_{GP}}(\hat L_Z;\r_0)$
grow by interaction-dependent values,
contrast Fig.~\ref{f3}c and Fig.~\ref{f3}a.
The combined effect is, of course, different than with $\r_0=(0.25,0,0)$,
and we discuss its main features,
focusing on the regime of interaction parameters larger than about $\Lambda=6$,
for which the two many-body curves ${\rm Var|_{MB}}(\hat L_X;\r_0)$ and ${\rm Var|_{MB}}(\hat L_Y;\r_0)$
cross, see Fig.~\ref{f3}c.
The many-body angular-momentum variances satisfy
${\rm Var|_{MB}}(\hat L_X;\r_0) > {\rm Var|_{MB}}(\hat L_Y;\r_0) > {\rm Var|_{MB}}(\hat L_Z;\r_0)$
for all studied interaction parameters.
So, the effect of this translation is to alter the 
order of the many-body variances,
i.e., to change the many-body anisotropy.
Now,
at the mean-field level,
we find 
${\rm Var|_{GP}}(\hat L_X;\r_0) > {\rm Var|_{GP}}(\hat L_Y;\r_0) > {\rm Var|_{GP}}(\hat L_Z;\r_0)$
up to about $\Lambda=12$
and
${\rm Var|_{GP}}(\hat L_X;\r_0) > {\rm Var|_{GP}}(\hat L_Z;\r_0) > {\rm Var|_{GP}}(\hat L_Y;\r_0)$
for the interaction parameters larger than about $\Lambda=12$.
All in all,
for the translation by
$\r_0=(0.25,0,0)$ the above-described relations
correspond, respectively, to 
the anisotropy classes $\{1\}$ and $\{1,2\}$
of the angular-momentum variances per particle.
 
Finally, 
for the translation by $\r_0=(0.25,0.25,0)$
non of the angular-momentum variances is invariant,
see (\ref{VAR_Lx_Ly_Lz_MB_TRANS_FINAL}) and (\ref{VAR_Lx_Ly_Lz_GP_TRANS_FINAL}).
The many-body angular-momentum variances 
${\rm Var|_{MB}}(\hat L_X;\r_0)$,
${\rm Var|_{MB}}(\hat L_Y;\r_0)$,
and
${\rm Var|_{MB}}(\hat L_Z;\r_0)$
are shifted by interaction-independent values
and
the mean-field angular-momentum quantities
${\rm Var|_{GP}}(\hat L_X;\r_0)$,
${\rm Var|_{GP}}(\hat L_Y;\r_0)$,
and
${\rm Var|_{GP}}(\hat L_Z;\r_0)$
increase by interaction-dependent values,
see Fig.~\ref{f3}d.
We examine the overall effect, 
concentrating on the regime of interaction parameters larger than about $\Lambda=2.5$,
where the two many-body curves ${\rm Var|_{MB}}(\hat L_X;\r_0)$ and ${\rm Var|_{MB}}(\hat L_Z;\r_0)$
cross, see Fig.~\ref{f3}d.
The many-body angular-momentum variances satisfy
${\rm Var|_{MB}}(\hat L_Y;\r_0) > {\rm Var|_{MB}}(\hat L_X;\r_0) > {\rm Var|_{MB}}(\hat L_Z;\r_0)$
for all interaction parameters in the range studied.
Once again,
the effect of the translation is to change the 
order of the many-body variances,
thereby altering the many-body anisotropy,
compare panels Fig.~\ref{f3}a,c,d.
At the mean-field level
one finds
${\rm Var|_{GP}}(\hat L_Y;\r_0) > {\rm Var|_{GP}}(\hat L_X;\r_0) > {\rm Var|_{GP}}(\hat L_Z;\r_0)$
up to about $\Lambda=15$,
then
${\rm Var|_{GP}}(\hat L_Y;\r_0) > {\rm Var|_{GP}}(\hat L_Z;\r_0) > {\rm Var|_{GP}}(\hat L_X;\r_0)$
till about $\Lambda=22.5$,
and
${\rm Var|_{GP}}(\hat L_Z;\r_0) > {\rm Var|_{GP}}(\hat L_Y;\r_0) > {\rm Var|_{GP}}(\hat L_X;\r_0)$
for all the interaction parameters larger than about $\Lambda=22.5$ studied,
see Fig.~\ref{f3}d.
Therefore,
for $\r_0=(0.25,0.25,0)$ the above-discussed findings
imply that all anisotropy classes can be attained
by the angular-momentum variances per particle
in the three-dimensional harmonic-interaction model at the infinite-particle-number limit.
These are, respectively,
$\{1\}$, $\{1,2\}$, $\{1,2,3\}$,
which rounds off the present work.

\section{Summary}\label{summary}

The present work deals with a connection between anisotropy of and correlations
in a three-dimensional trapped Bose-Einstein condensate.
The merit of treating the limit of an infinite number of bosons is appealing,
since the system is known to be 100\% condensed in this limit,
and some of its properties,
notably the density per particle,
are identical at the many-body and mean-field levels of theory.

We have analyzed the variances per particle of the three Cartesian components of the 
position ($\hat X, \hat Y, \hat Z$),
momentum ($\hat P_X, \hat P_Y, \hat P_Z$),
and angular-momentum ($\hat L_X, \hat L_Y, \hat L_Z$) operators
at the many-body and mean-field levels of theory.
In generally, for small interaction parameters
the differences between the many-body and mean-field quantities
are quantitative
whereas for larger interaction parameters qualitative differences emerges.
We define the anisotropy class of the variance according
to the different orderings from large to small,
or permutations,
of the three respective many-body and mean-field quantities.
The anisotropy class $\{1\}$ implies the same ordering,
the class $\{1,2\}$ implies that two of the components are permuted,
and the anisotropy class $\{1,2,3\}$ means that the three components
are permuted.

Two relatively transparent applications are presented,
the first is the breathing of an anisotropic three-dimensional trapped Bose-Einstein condensate,
and the second is the ground state of the anisotropic three-dimensional harmonic-interaction model.
The former exhibits different anisotropy classes of the position and momentum variances per particle,
because at the many-body level
the variances are constant,
while at the mean-field level
the variance of each component oscillates with a different amplitude and frequency,
and consequently different anisotropy classes occur in time.
The latter application shows different anisotropy classes of the angular-momentum variance per particle,
owing to the intricate transformation properties of the angular-momentum variance
when the wave-packet is translated from the origin.

To sum up, the anisotropy or morphology of a three-dimensional trapped Bose-Einstein condensate
can look quite different when examined through the `glasses'
of many-body and mean-field theories,
even for 100\% condensed bosons at the limit of an infinite number of particles.
It would be interesting to conduct the investigation presented here
and analyze the results 
in more complicated numerical many-particle scenarios.
Last but not least,
it is possible to envision classifying the morphology of a Bose-Einstein
condensate beyond that emanating from the variances
of the position, momentum, and angular-momentum operators.

\section*{Acknowledgements}

This research was supported by the Israel Science Foundation (Grants No. 600/15 and 1516/19).
We thank Anal Bhowmik and Lorenz Cederbaum for discussions.
Computation time on the BwForCluster,
the High Performance Computing system Hive of the Faculty of Natural Sciences at University of Haifa,
and the Hawk at the High Performance Computing Center
Stuttgart (HLRS) is gratefully acknowledged.

\appendix

\section{Translated angular-momentum variances in three spatial dimensions at the
limit of an infinite number of particles}\label{APP}

Consider interacting bosons at the limit of an infinite number of particles trapped in the ground state
of the three-dimensional anisotropic harmonic potential
(or any other potential which is reflection symmetric in the $x$, $y$, and $z$ directions) centered at the origin.
The expectation values per particle of the position, momentum, and angular-momentum operators vanish.
Suppose now that the harmonic trap is translated to the location $\r_0=(x_0,y_0,z_0)$. 
The translated and untranslated angular-momentum variances at the many-body and mean-field levels
of theory can, respectively, be related as follows:
\beqs\label{VAR_Lx_Ly_Lz_TRANS}
\beqn\label{VAR_Lx_Ly_Lz_MB_TRANS}
& &
{\rm Var|_{MB}}(\hat L_X;\r_0) = {\rm Var|_{MB}}(\hat L_X) + y_0^2 {\rm Var|_{MB}}(\hat P_Z) + z_0^2 {\rm Var|_{MB}}(\hat P_Y),
\nonumber \\
& & 
{\rm Var|_{MB}}(\hat L_Y;\r_0) = {\rm Var|_{MB}}(\hat L_Y) + z_0^2 {\rm Var|_{MB}}(\hat P_X) + x_0^2 {\rm Var|_{MB}}(\hat P_Z),
\nonumber \\
& &
{\rm Var|_{MB}}(\hat L_Z;\r_0) = {\rm Var|_{MB}}(\hat L_Z) + x_0^2 {\rm Var|_{MB}}(\hat P_Y) + y_0^2 {\rm Var|_{MB}}(\hat P_X)
\
\eeqn
and
\beqn\label{VAR_Lx_Ly_Lz_GP_TRANS}
& &
{\rm Var|_{GP}}(\hat L_X;\r_0) = {\rm Var|_{GP}}(\hat L_X) + y_0^2 {\rm Var|_{GP}}(\hat P_Z) + z_0^2 {\rm Var|_{GP}}(\hat P_Y),
\nonumber \\
& &
{\rm Var|_{GP}}(\hat L_Y;\r_0) = {\rm Var|_{GP}}(\hat L_Y) + z_0^2 {\rm Var|_{GP}}(\hat P_X) + x_0^2 {\rm Var|_{GP}}(\hat P_Z),
\nonumber \\
& &
{\rm Var|_{GP}}(\hat L_Z;\r_0) = {\rm Var|_{GP}}(\hat L_Z) + x_0^2 {\rm Var|_{GP}}(\hat P_Y) + y_0^2 {\rm Var|_{GP}}(\hat P_X).
\
\eeqn
\eeqs
The explicit expressions at the many-body and mean-field levels of theory
are given in the main text, see (\ref{VAR_Lx_Ly_Lz_MB_TRANS_FINAL})
and (\ref{VAR_Lx_Ly_Lz_GP_TRANS_FINAL}), respectively.
We remind for reference that the position and momentum variances are invariant to translations.
Consequently, by subtracting the Gross-Piteavskii (\ref{VAR_Lx_Ly_Lz_GP_TRANS}) from the
many-body (\ref{VAR_Lx_Ly_Lz_MB_TRANS}) results,
see (\ref{VAR_INF}),
we readily find for the translated
correlations terms:
\beqn\label{VAR_Lx_Ly_Lz_CORR_TRANS}
& &
{\rm Var|_{correlations}}(\hat L_X;\r_0) =
{\rm Var|_{correlations}}(\hat L_X) + y_0^2 {\rm Var|_{correlations}}(\hat P_Z) + z_0^2 {\rm Var|_{correlations}}(\hat P_Y),
\nonumber \\
& &
{\rm Var|_{correlations}}(\hat L_Y;\r_0) =
{\rm Var|_{correlations}}(\hat L_Y) + z_0^2 {\rm Var|_{correlations}}(\hat P_X) + x_0^2 {\rm Var|_{correlations}}(\hat P_Z),
\nonumber \\
& &
{\rm Var|_{correlations}}(\hat L_Z;\r_0) =
{\rm Var|_{correlations}}(\hat L_Z) + x_0^2 {\rm Var|_{correlations}}(\hat P_Y) + y_0^2 {\rm Var|_{correlations}}(\hat P_X).
\nonumber \\ \
\eeqn
The meaning of result (\ref{VAR_Lx_Ly_Lz_CORR_TRANS}) is that
the correlations terms of the translated angular-momentum variances
depend on the respective translated correlations terms of the momentum variances
and components of the translation vector $\r_0$.
The translated correlations terms (\ref{VAR_Lx_Ly_Lz_CORR_TRANS}) generally
depend more strongly on the interaction parameter
than the untranslated ones.
Indeed, Fig.~\ref{f3} plots some examples of angular-momentum variances for different $\r_0$,
and shows that,
once translations of the trap are included,
the angular-momentum variances
of a trapped Bose-Einstein condensate at
the limit of an infinite number of particles 
can belong to any of the anisotropy classes,
$\{1\}$, $\{1,2\}$, or $\{1,2,3\}$.

\end{document}